\documentclass[%
 aip,
prb,%
 amsmath,amssymb,
reprint,%
showpacs,
]{revtex4-1}

\usepackage{graphicx}
\usepackage{dcolumn}
\usepackage{bm}
\usepackage{blkarray}
\usepackage[usenames,dvipsnames]{color}
\usepackage[hyperfootnotes=false]{hyperref}
\usepackage[hyperfootnotes=false]{hyperref}
\usepackage{threeparttable}
\hypersetup{
 colorlinks,
 citecolor=Violet,
 linkcolor=Red,
 filecolor= green,
 urlcolor=Blue}
\begin{document}

\title{Enhancement of thermoelectric figure-of-merit of Graphene upon BN-doping and sample length reduction}

\author{Ransell D'Souza}%
\email{ransell.d@gmail.com; ransell.dsouza@bose.res.in}

\author{Sugata Mukherjee}%
\email{sugata@bose.res.in; sugatamukh@gmail.com}

\affiliation{Department of Condensed Matter Physics and Materials Science, S.N. Bose National Centre for Basic Sciences, Block JD, Sector III, Salt Lake, Kolkata 700106, India}

\begin{abstract}
Using first-principles density functional perturbation theory based calculations of length-dependent lattice thermal conductivity ($\kappa_L$) and using our previously calculated results (Phys Rev B {\bf95} 085435 (2017)) of electrical transport, we report results of thermoelectric figure-of-merit ($ZT$) of monolayer and bilayer Graphene. We find nearly ten-fold increase in $ZT$ for the graphene sample doped with boron nitride and reduced sample length. We also compare $\kappa_L$ calculated using the iterative real space method with conventional analytical Callaway-Klemens method and obtain the flexural (ZA) phonon modes to be dominant in thermal transport unlike in the latter method.
Our calculations are in good agreement with available experimental data.
\end{abstract}



\maketitle

\section{Introduction}
Graphene is single-atom thick, $sp^2$ hybridized carbon atoms arranged in a two-dimensional (2D) honeycomb
 crystal structure having two atoms in its unit cell \cite{geim07}.
 Bilayer graphene consists of two monolayers, with four atoms in the unit cell arranged in an AB-type stacking known as Bernal-stacked.
 Monolayer graphene (MLG), has a linear band dispersion and is a semi-metal with a high electrical mobility \cite{kim08, kim09, novoselov05, zhang05, schedin07, tan07, chen08}.
 On the other hand bilayer graphene (BLG) can be used as a tunable band gap semiconductor \cite{hongki07}.
 These intriguing properties along with the fact that MLG and BLG are extremely atomically
  stable make them ideal candidates for a good testing ground since they can be supported between two leads.
Being a planar 2D structure, one of the main advantage of MLG and BLG over many other materials is that they
can be readily used in circuit designs with standard lithography methods and hence can be used to fabricate transistors at true nanoscale limit.
In the interest of thermoelectric devices and fundamental physics, MLG and BLG have therefore drawn a large extent of experimental, computational and theoretical attention for their transport properties.

Graphene and other related 2D nanomaterials exhibit crucial important properties which are useful for their application in renewable energy \cite{repp18,pham16}. Graphene nanoribbons are found to be promising candidates for power generation in thermoelectric devices. Experimentally it has been shown that the thermal transport of graphene based devices can be tailored by defects, isotope engineering, edge roughness and by techniques to introduce nano-holes \cite{Bonaccorso15}. From a theoretical point of view, study of the lattice thermal conductivity of MLG and BLG is of great significance \cite{lindsay14,saito18}. 

 The quality for being able to efficiently generate thermoelectric power in transport devices is related to the dimensionless quantity figure-of-merit ($ZT$), expressed by $\frac{S^2\sigma T}{\kappa}$. Where, $S$ is the Seebeck coefficient, $\sigma$ is the electrical conductivity, $T$ is the temperature and $\kappa$ is total (contribution due to electrons and the lattice) thermal conductivity. An efficient material will hence require a high power factor ($S^2\sigma$) and low thermal conductivity.
 We have recently \cite{RDSM17} showed that BN-doped MLG and BLG exhibit increase in the power factor, as such doping induces a small band gap, thereby increasing $S$.
 Therefore calculations on thermal conductivity of these materials are highly desirable because if the thermal conductivity decreases, doping or inducing graphene with impurities would be a useful technique to increase the power performance of graphene devices.

 The relative contributions to heat conduction by the acoustic in-plane and out-of-plane phonons are still debatable. Several studies \cite{kuang16,lindsay11,lindsay2011,seol10,lindsay2010} indicate that out-of-plane ZA phonon modes to be the most dominant while few others \cite{shen14,alofi13,kong2009,aksamija11,nika2009,nika11,nika12,wei14} report the opposite. Therefore an analytical expression for the acoustic modes would play a very important role in solving these discrepancies. In this paper, we attempt to resolve such issues. 
 
Allen has shown that the Callaway method \cite{callaway59} underestimates the suppression of the normal processes and has proposed an improved method \cite{allen13} which has been compared with the iterative method before for various materials \cite{ma14}. However, this has been done only for three dimensional materials.
There have also been earlier studies on the length dependence thermal conductivity of single layer graphene using either a Monte Carlo simulation on a quadratic and linear fit to the acoustic phonon modes \cite{mei14} or the improved Callaway model \cite{majee16,guo17}. 

In all of these calculations, none of the relaxation times were calculated beyond the relaxation time approximation (RTA) using an iterative method. 
All length dependence calculations were done at room temperature.
Moreover, those calculations do not take the symmetry of the sample into account. 

Previous studies using density functional perturbation theory (DFPT)to calculate the thermal conductivity by solving the BTE exactly have been reported for graphene \cite{lindsay14}, bilayer graphene \cite{fugallo14}, and, B and N doped graphene \cite{polanco18}.
For example, Fugallo {\it et al.}\cite{fugallo14} have solved the BTE exactly for MLG and BLG using phonon-phonon scattering rates derived from DFPT. Their calculations concentrate on the collective phonon excitations in comparison to the single phonon excitations. Our calculations, on the other hand, deal with solving the BTE beyond the RTA for each acoustic mode at various lengths and temperatures.
The temperature dependent behavior of each acoustic mode using the iterative method has been demonstrated by Lindsay {\it et. al} \cite{lindsay14} only for MLG. 

Previous theoretical calculations on mode dependent lattice thermal conductivity for BLG have used the Tersoff potential \cite{lindsay2011} and hence a first-principle calculation showing the length, temperature dependence of the lattice thermal conductivity is of great importance.
Concentration dependent lattice thermal conductivity for B and N doped graphene calculations suggest a decrease in lattice thermal conductivity \cite{polanco18}, a feature similar to what is seen in our calculations.
Reproducing these previously reported results for graphene justifies our predictions for the temperature dependent lattice thermal conductivity for BLG.
 Merging the lattice thermal conductivity calculations, which are in good agreement with previous theoretical and experimental data, with the calculations of the electrical transport parameters ensure that our calculated figure of merit is accurate for MLG, BLG, and BN doped MLG.

In this paper we provide an analytical solution using the Callaway-Klemens method using a quadratic fit to the out-of-plane ZA and linear fit to the in-plane LA and TA acoustic phonon dispersion, respectively, and show that the results are in perfect agreement with experiments for MLG at room temperature only. The Callaway-Klemen method overestimates $\kappa_L$ at lower temperatures for MLG and overestimates $\kappa_L$ at all temperatures and lengths for BLG when compared to experimental measurements. This motivated the present study on the thermal conductivity by calculating the scattering rates for each acoustic mode beyond the RTA using the first-principles iterative ShengBTE \cite{ShengBTE} method for MLG and BLG at various lengths as well as temperatures and compare our results with various experimental data.
Moreover, using the scattering rates from the  ShengBTE method, we also calculate the length dependent $\kappa_L$ for doped MLG treating BN dimers as point defects.

Using the above accurate calculations of $\kappa_L$ and 
our earlier calculated data \cite{RDSM17} on $S$ and $\sigma$ for these materials, we calculate the thermoelectric figure-of-merit ($ZT$) of MLG and BLG and study the effect of sample length and BN-doping on $ZT$. Purpose of this paper is to demonstrate that the iterative ShengBTE method gives accurate results of $\kappa_L$ and its dependence on sample length for both MLG and BLG without any fitting parameters in quantitative agreement with experimental data. We have also found $\kappa_L$ to  decrease ($\sim 70\%$) for MLG upon BN-doping. Our calculated decrease in $\kappa_L$ is consistent with such decrease observed in oxygen defects in Graphene using Raman spectroscopy \cite{anno17}. 
We have presented a comparison of our results with experimental data using the local activation model \cite{jorio11}.
This decrease in $\kappa_L$ together with increase in $S$ and $\sigma$ leads to nearly ten fold increase in the thermoelectric figure-of-merit ($ZT$) upon BN-doping and decrease in sample length.

In the next section our calculational method is presented together with the theoretical framework used for the calculation of lattice thermal conductivity $\kappa_L$. Results on the phonon dispersion, Gr\"uneisen parameter, $\kappa_L$ and eventually $ZT$ are given in subsequent sections followed by a summary.

\section{Method of Calculation}
\subsection{Electronic and phonon band dispersion} 
Geometry optimizations were carried out on a hexagonal unit cell for both monolayer (MLG) and bilayer graphene (BLG) using the first-principles Density functional theory (DFT) as implemented in the QUANTUM ESPRESSO code \cite{giannozzi09}. The unit cell consists of 2 and 4 carbon atoms in the $xy$-plane for MLG and BLG, respectively.
Ultrasoft pseudopotential was used to describe the exchange-correlation potential kernel in the local density approximation (LDA) \cite{rrkj90}.
A vacuum spacing of 20 \AA \ was introduced in the $z$-direction for the periodic supercell, which avoid interactions between atoms in different planes for MLG. For BLG two such layers were taken in such supercell where Van der Waals interaction \cite{grimme1} between the planes was included. For the $k$-point sampling, we have chosen a Monkhorst-Pack \cite{mp76} grid of $16 \times 16 \times 1$ and $16 \times 16 \times 4$ for MLG and BLG, respectively. A 160 Ry charge density energy cut-off and 40 Ry kinetic energy cut-off were used in solving the Kohn-Sham equation self consistently with an accuracy of 10$^{-9}$ Ry. 

 The phonon dispersion along the high-symmetry points in the two-dimensional (2D) hexagonal Brillouin zone (BZ) ($q_z=0$), and phonon density of states (PDOS) were calculated on the geometrically optimized structures
  using the density functional perturbation theory (DFPT) \cite{dfpt87}.
  A Monkhorst-Pack $q$-grid of $6 \times 6 \times 1$ for MLG and $6 \times 6 \times 2$ for BLG was used in the
  self-consistent calculations with a phonon frequency threshold of 10$^{-14}$ cm$^{-1}$.

\subsection{Theoretical methods for calculation of $\kappa_L$}
\subsubsection{Analytical Callaway-Klemens method}
The lattice thermal conductivity ($\kappa_L$) by Callaway-Klemens \cite{klemens58,callaway59} method and modified by Nika {\it et. al.} \cite{nika2009} for an isotropic two-dimensional system is expressed as
\begin{eqnarray} \label{k}
\kappa_L &=& {1\over 4\pi k_B T^2 N \delta}  \nonumber \\
 &\times& {\sum\limits_{s} \int\limits_{q_{min}}^{q_{max}}[\hbar \omega_s(q)]^2 v_s^2(q) \tau_{tot}(q)\frac{e^{\frac{\hbar \omega_s(q)}{k_B T}}}{[e^{\frac{\hbar \omega_s(q)}{k_B T}}-1]^2} q dq},
\end{eqnarray}
where, $\hbar$ is the reduced Planck constant, $\delta$ is the hight between two consecutive layers, $N$ number of layers, $k_B$ is the Boltzmann constant. 
The mode dependent phonon frequency and velocity at wave vector $q$ and corresponding to branch $s$ is denoted by $\omega_s(q)$ and $v_s(q)$.
$q_{max}$ is wave vector corresponding to the Debye frequency while $q_{min}$ is the wave-vector that
 corresponds to the sample length ($L$) dependent low cut-off frequency. We have used this formulation to calculate $\kappa_L$ for 2D single- and multilayered BN
and its length dependence previously \cite{RSBN17}.

The total phonon relaxation time $\tau_{tot}$ comprises of the contributions from,
(i) the phonon-phonon Umklapp scattering, (ii) the boundary scattering, and (iii) scattering due to point defects.
The phonon-phonon Umklapp scattering rate ($\tau_U$) for a given mode $s$ is given by \cite{klemens58,callaway59,nika2009}
\begin{eqnarray}\label{tauU}
\tau_{U,s}(q) = \frac{Mv_s^2(q)\omega_{D,s}}{\gamma_s^2(q) k_B T \omega_s(q)^2},
\end{eqnarray}
where, $M$ is the total mass of the atoms in the unit cell and $\gamma_s(q)$ is the mode dependent Gr\"{u}neisen parameter at wave vector $q$.
The Gr\"{u}neisen parameter ($\gamma_s(q)$) was calculated by us \cite{RDSM17} recently by applying a $\pm$ 0.5\% biaxial strain to MLG and BLG using $\gamma_s(q) = \frac{-a_0}{2\omega_s(q)}\frac{\delta\omega_s(q)}{\delta a}$.

The rough boundary scattering rate are shown to be given by \cite{nika2009},
\begin{eqnarray}\label{tauB}
\tau_{B,s}(q) = \frac{d}{v_s\big(\omega_s(q)\big)}\frac{1+p}{1-p},
\end{eqnarray}
where, $d$ is the width of the sample. The specularity parameter ($p$) depends on the roughness of the edges. For example an ideal smooth sample would have a specularity parameter $p=1$.
The scattering rate due to point defects is written as\cite{nika2009},
\begin{eqnarray}\label{tauP}
\tau_{P,s}(q) = \frac{4v_s\big(\omega_s(q)\big)}{S_0\Gamma_0 q_s\big(\omega_s(q)\big)}\frac{1}{\omega(q)_s^2},
\end{eqnarray}
where, $\Gamma_0$ is a dimensionless parameter to determine the strength of the point-defect scattering, given by $\Gamma_0 = \sum_{i}f_i(1-\frac{M_i}{\overline M})$, where $\overline{M} = \sum_i M_i f_i$ is the average atomic mass, $f_i$ is the fractional concentration of the impurity atoms with mass $M_i$. The cross-sectional area per atom of the lattice is denoted by $S_0$.
Each of the mentioned scattering rates can be combined to calculate the total phonon relaxation time which is given by the Matthiessen's rule \cite{ashcroft},
\begin{eqnarray}\label{tau}
\frac{1}{\tau_{tot,s}(q)} = \frac{1}{\tau_{U,s}(q)} + \frac{1}{\tau_{B,s}(q)} + \frac{1}{\tau_{P,s}(q)}.
\end{eqnarray}

The integral in Eq.\ref{k} over the entire BZ would require setting the lower cut-off to zero which would lead to an infinite
 thermal conductivity. Klemens \cite{klemens94} offered a physical reason for selecting the cut-off frequency corresponding to the ZO$'$ mode at the $\Gamma$ point for bulk graphite. This method would work with BLG but would not for MLG due to the absence of the ZO$'$ mode.
 
 Another approach to avoid the divergence is by curbing the phonon mean free path (MFP) on the boundaries of the sheets \cite{nika11}. 
This is achieved by making a condition that the MFP cannot exceed the physical size of the sample.
 This has been incorporated in both the approaches in the present paper.

By changing the integral variable from wave-vector ($q$) to phonon frequency ($\omega$), the upper and lower limits of the integral in Eq. \ref{k} would then correspond to the Debye ($\omega_D$) and cut-off ($\omega_{min})$ frequency, respectively, which has been described in a recent paper \cite{RSBN17}. 

A simple analytical expression for the mode-dependent $\kappa_L$ using the Callaway-Klemens method with point defects and boundary scattering is given in the appendix.


\subsubsection{Iterative real-space method (ShengBTE)}
We have also used an alternative first-principles iterative method based on DFT and DFPT utilizing third-order anharmonic inter-atomic force constants for calculating $\kappa_L$ of MLG and BLG, as implemented in ShengBTE code \cite{ShengBTE}. In this method
the lattice thermal conductivity tensor $\kappa_{L}^{\alpha \beta}$ is calculated by solving the phonon Boltzmann transport equation from the converged set of phonon scattering rates using the expression, 
{\small \begin{eqnarray}\label{kl}
\kappa_L^{\alpha \beta}=\frac{1}{k_BT^2 \Omega\,N}\sum_sf_0(f_0+1)(\hbar \omega_s)^2v_s^{\alpha} \tau_{tot,s} (v_s^{\beta}+\Delta_s^{\beta}).
\end{eqnarray}
Here, $\Omega$ is the volume of the unit-cell, $N$ denotes the number of $q$-points in the BZ sampling. Here, $\tau_{tot}$, $\omega$ and $v$ are the total phonon relaxation time, frequency and the phonon group velocity, respectively. $f_0$ denotes the Bose-Einstein distribution function and $\Delta$ accounts for the correction to the phonon group velocity due to the deviation from the relaxation time approximation \cite{ShengBTE,RSBN17}. It must be noted that in the iterative {\it ab-initio} method, the scattering rate processes can not be classified into Umklapp or Normal scattering processes as done in the Callaway-Klemens method (Eq. \ref{tauU}). The solutions to the self consistent equations are the combined three-phonon scattering rates \cite{ShengBTE}.

Lattice thermal conductivity calculations for MLG and BLG using ShengBTE method, having two and four atoms in its unit cell, using third nearest neighbor interactions for a $4 \times 4 \times 2$ supercell yields 72 and 156 displaced supercell configurations, respectively. From a set of third order derivatives of energy, calculated by implementing the plane-wave method on these displaced supercell configurations, the third-order anharmonics IFCs are constructed.
The number of such configurations increase exponentially with increase in number of atoms in the unit cell.
For example, five layered boron nitride, having 20 atoms in the unit cell, the same number of nearest neighbor interaction for the calculation of $\kappa_L$ required 828 configurations as used earlier \cite{RSBN17}.

Finally, from our previous results \cite{RDSM17} of the electrical conductivity ($\sigma$) and the Seebeck coefficient ($S$), we obtain the figure-of-merit ($ZT$) of undoped MLG and BLG and BN-doped MLG, using $ZT = \frac{S^2 \sigma\, T}{\kappa_{el} + \kappa_L}$, where $\kappa_{el}$ is the electronic thermal conductivity, found to be $\sim10^{-8}$ times smaller than $\kappa_L$.

\section{Results and discussions}
\subsection{Phonon dispersion, density of states and the Gr\"uneisen parameters}
%

In Fig. \ref{phdos} we show the calculated phonon dispersion along the high symmetry $q$-points in the 2D irreducible hexagonal BZ for MLG and BLG \cite{RDSM17} and the phonon densities of states (PDOS). 
Using standard group theoretical methods, it can be shown \cite{RSBN17} that for MLG the six allowed phonon modes are represented by the point-group (PG) symmetry $A_{2u}$ + $E_{2g}$ + $B_{1g}$ + $E_{1u}$, whereas for BLG the twelve allowed modes are represnted by the PG symmetry  2$A_{2u}$ + 2$E_{g}$ + 2$A_{g}$ + 2$E_{u}$.
Transitions corresponding to the basis $x,y,z$ are Infrared active while transitions corresponding to product of those basis ($xy, yz, x^2$) are Raman active.
The momentum conservation requires that the first-order Raman scattering processes are limited to the phonons at the center of BZ ($q = 0$).
Calculated allowed Raman and Infrared phonon frequencies for MLG and BLG and available experimental data are shown in table \ref{IR}.

\begin{figure}[!htbp]
\centering \includegraphics[scale=0.35]{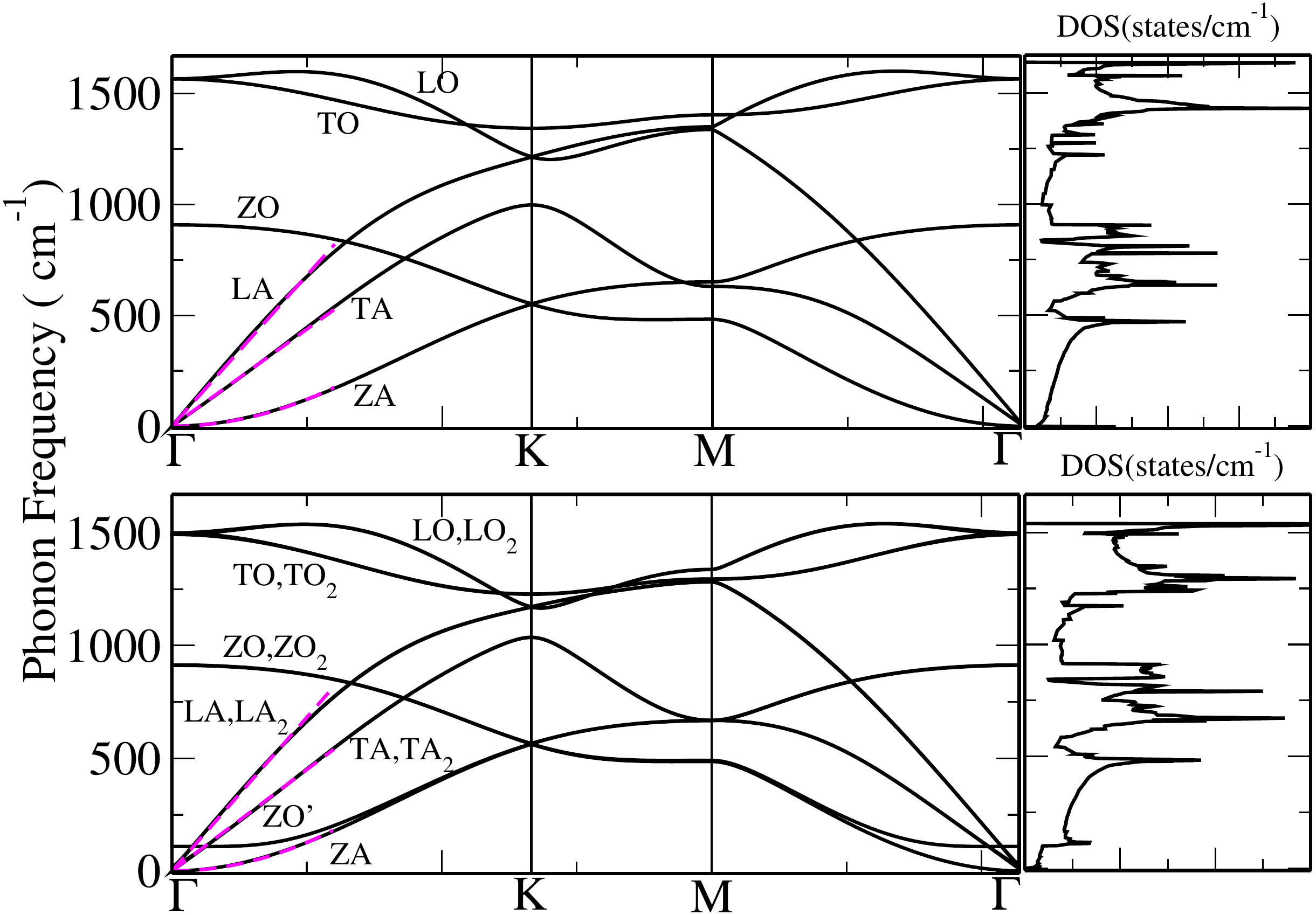}
\caption{\label{phdos} Calculated phonon dispersion and phonon density of states of MLG (above) and BLG (below) along the high symmetry points of the 2D hexagonal Brillouin zone. The magenta dashed lines are the best linear and quadratic fit to the in-plane and out-of-plane wave dependent fit to the phonon dispersion. }
\end{figure}

We have used linear fit to the in-plane acoustic phonon modes (LA, TA) and a quadratic fit to the out-of-plane acoustic mode (ZA), for obtaining an analytical form of mode-dependent $\kappa_L$ for these materials as discussed in the Appendix.

\begin{table}[t]
\centering
\caption{\label{IR} Phonon frequencies at the $\Gamma$ point derived from our earlier calculations in comparison with experimentally measured Raman frequencies.}
\begin{tabular}{c|c|c|c}
\hline\hline
$\omega$(cm$^{-1}$)  & Expt. (Sys.) & MLG (P.G. Sym.) & BLG (P.G. Sym.) \\
  \hline
$\Gamma_{\rm ZO}$ & 867.9 (MLG$^{\rm a}$) & 907 ($A_{2u}$) & 915 ($A_{2u}$) \\
$\Gamma_{\rm LO}$ & 1579.7(MLG$^{\rm a}$) & 1580.0 ($E_{2g}$) & 1544 ($E_{g}$) \\
$\Gamma_{\rm TO}$ & 1579.7 (MLG$^{\rm a}$) & 1580.0 ($E_{1u}$) & 1540 ($E_{u}$) \\
$\Gamma_{\rm ZO^{'}}$ & 99 (BLG$^{\rm b}$) & - & 108 ($A_{1g}$)  \\
$\Gamma_{\rm LA_{2}}$ & 32 (BLG$^{\rm c}$) & - & 22.16 ($E_{g}$)  \\
$\Gamma_{\rm TA_{2}}$ & 32 (BLG$^{\rm c}$) & - & 22.16 ($E_{g}$)  \\
\hline
$\gamma$  & Expt (Sys.) & MLG & BLG \\
  \hline
  $\gamma_{\rm TO}$ & 1.99 (MLG$^{\rm d}$) & 1.85 & 1.89 \\
   \hline \hline
\end{tabular}
\begin{tablenotes}
\item
$^{\rm a}$ Experimental data, Ref. \cite{dresselhaus09,yanagisawa05}. \\
$^{\rm b}$ Experimental data derived from overtone Raman peaks, Ref. \cite{lui13}. \\
$^{\rm c}$ Experimental Raman data, Ref. \cite{tan12}. \\
$^{\rm d}$ Experimental Raman data, Ref. \cite{mohiuddin09}. \\
\end{tablenotes}
\end{table}

As in the case of the first and second order Raman and Infrared spectroscopy \cite{lui13,tan12}, where
our calculations based on harmonic IFCs are justified experimentally, the measurements
based on the pressure dependence of Raman lines \cite{mohiuddin09} sheds light on the anharmonic IFCs. 
The mode dependent Gr\"{u}neisen parameters ($\gamma$) can be obtained experimentally at the high-symmetry $\Gamma$ point using the mentioned technique\cite{mohiuddin09}.

Gr\"{u}neisen parameters along the $\Gamma$ to K direction for MLG and BLG are plotted in Fig. \ref{gp}.
Alongside the first-principles Gr\"{u}neisen parameter calculations, we plot the fitted constant values that we use in our study to calculate the contributions to $\kappa_L$ from the in-plane acoustic phonon modes (LA, TA) and the fitted inverse square dependence on the wave-vector ($q$) for the out-of-plane acoustic mode (ZA).

The inverse wave vector squared ($1/q^2$) dependence of $\gamma$ for the ZA mode can be easily understood from the definition of Gr\"{u}neisen parameter for two-dimensional crystal \cite{RSBN17}, $\gamma_s(q) = -(a_0/2\omega_s(q))(\delta \omega_s(q)/\delta a)$, and from the $q^2$-dependence of the phonon dispersion of the ZA mode.
Here, $a_0$ and $\delta a$ denote the lattice constant and its change under strain, respectively.
Under small positive and negative strain $\gamma_{ZA}(q) \sim 1/q^2$, since the second term in $\gamma_{ZA}(q)$ will not depend on $q$.

\begin{figure}[h]
\centering \includegraphics[scale=0.32]{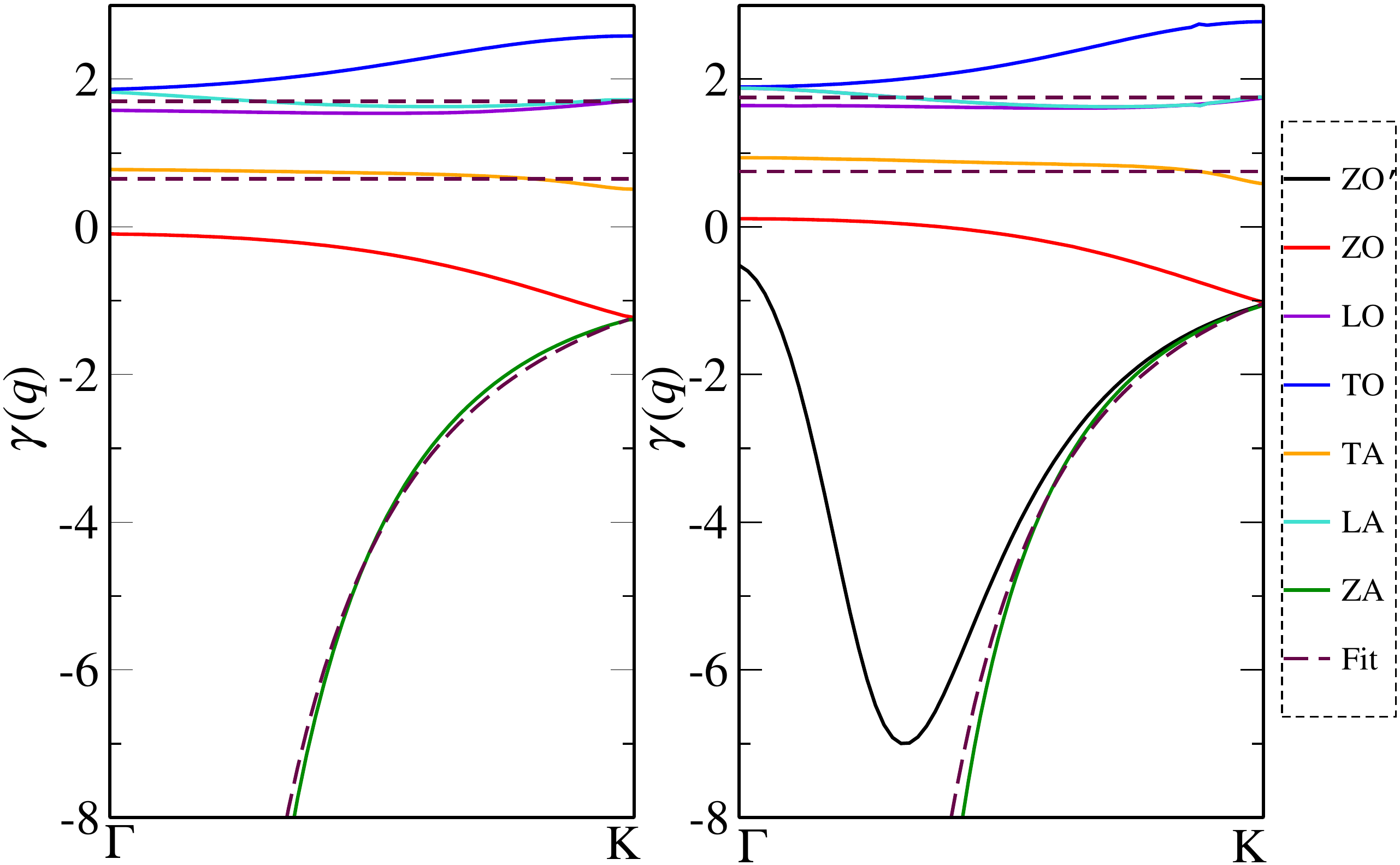}
\caption{\label{gp}The calculated Gr\"{u}neisen ($\gamma$) parameters of all the modes along the $\Gamma$ to K direction of the 2D Brillouin zone of the hexagonal unit cell for MLG(left) and BLG(right). The maroon dashed lines are the best constant and inverse squared wave dependent fits to the in-plane (LA,TA) and out-of-plane (ZA) $\gamma$ parameters, respectively.}
\end{figure}

It is clear from the figure that the $\gamma$ parameters for the in-plane modes (LA,TA) do not deviate much from their
average value justifying a constant approximation made in our calculations.
Similarly, the fit to the $\gamma$ parameters using 
 the fitted inverse square wave dependence for the out-of-plane ZA mode is in good agreement with our first-principle calculations.
The $\gamma$ parameters provide information on the degree phonon scattering and anharmonic interactions between lattice waves \cite{marzari05}. From Eq. \ref{tauU} it is clear that the relaxation time is highly dependent on $\gamma$ which in turn is dependent on $\kappa_L$. 

In the next section we calculate, using the phonon dispersion and the Gr\"{u}nesien parameters, the lattice thermal conductivity with and without point defects.

\subsection{Lattice thermal conductivity using the Callaway-Klemens method}
We first obtain the analytical solution of $\kappa_L$ for an ideal sheet of MLG and BLG and then calculate $\kappa_L$ numerically for them with defects and specularity parameter ($p$) with values other than one.

\begin{figure}[!htbp]
\centering \includegraphics[scale=0.32]{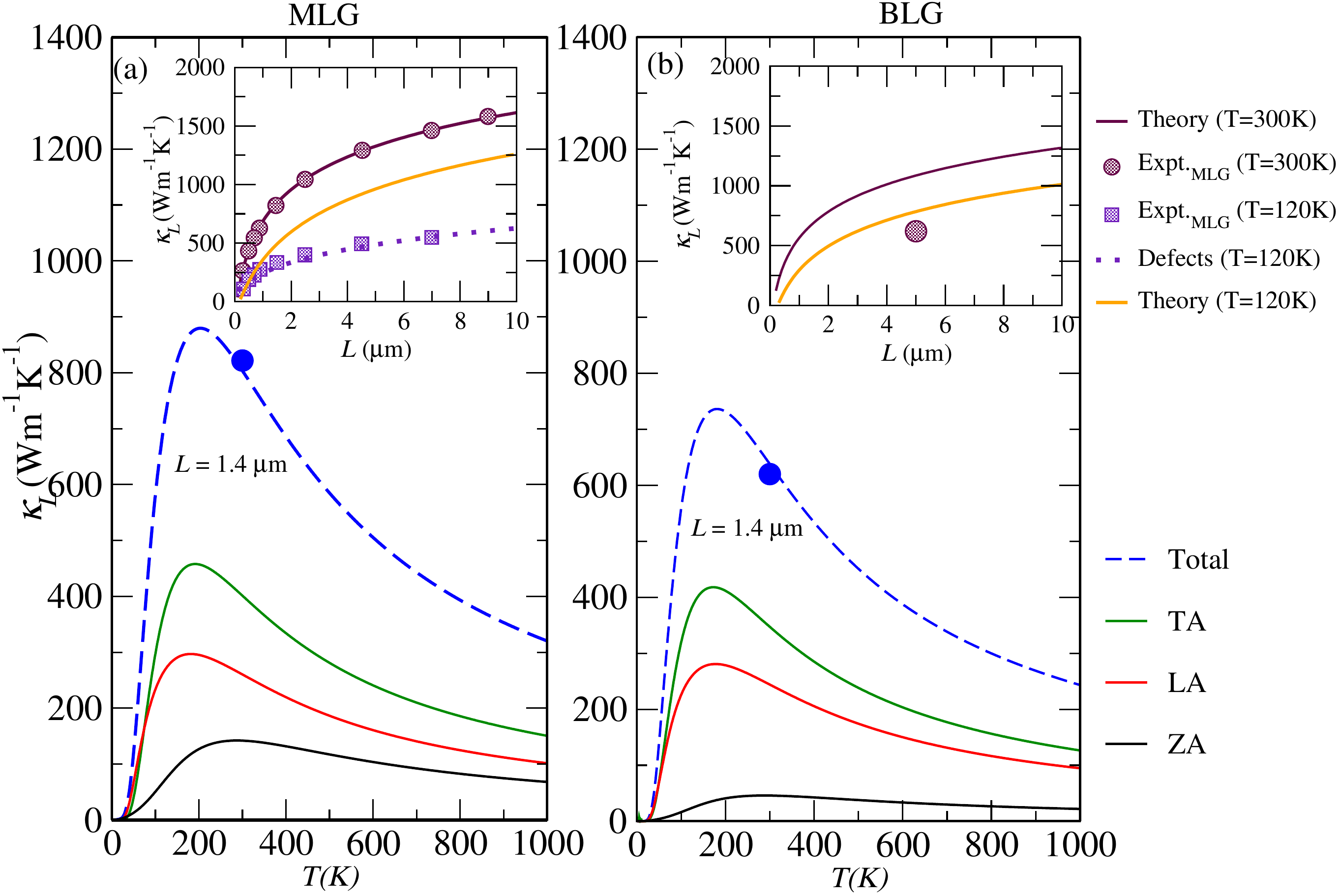}
\caption{\label{kfig} Temperature dependence of $\kappa_L$ using the analytical solutions of the Callaway-Klemens method for each of the acoustic modes at a constant length for (a) MLG and (b) BLG. The blue dots are experimental values of $\kappa_L$ at room temperature for sample lengths 1.4$\mu$m and 5$\mu$m for MLG and BLG, respectively. 
Inset: Length dependence of $\kappa_L$ at constant temperatures, $T$=120K and $T$=300K. 
The maroon dotted lines are the length dependence with point defects with parameters used to fit the experimental data \cite{pettes11,xu14}. }
\end{figure}

Fig. \ref{kfig} shows the acoustic mode dependent $\kappa_L$ as a function of temperature for (a) MLG and (b) BLG at a constant length of 1.4 $\mu$m. 
The insets in Fig. \ref{kfig} (a) and (b) are the length dependent $\kappa_L$ at two constant temperatures 120K and 300K. 
Length is defined as the direction along which the heat propagates.
The fitting parameters, discussed in the previous section, used in this study are shown in table \ref{para}.
The length dependent analytical form for $\kappa_L$ 
for MLG at RT is in excellent agreement with a recent experiment \cite{xu14}. However, our results
 of $\kappa_L$ overestimate the experimental data at $T$=120K.
 One explanation for this could be that in the lower temperature range (0-100 K), $\kappa_L$ increases rapidly and hence a small change in the temperature in this range would result
 into a large change in the lattice thermal conductivity, making comparison with experimental data difficult.
 Another possible reason could be that we have considered a sample with an ideal sheet without any form of defects or impurity.
 We find that, using a large specularity parameter of $p=0.9$ and an extremely small value of
  $\Gamma_0$=0.001 of Eq. \ref{tauP}, our length dependent calculations with point defects for MLG agree with experimental measurements at $T$=120K. 

The major difference between the thermal conductivity of MLG and BLG is due the out-of-plane ZA phonon mode.
MLG has a total of twelve process involving the flexural phonons (ZA). 
Seol {\it et. al.} \cite{seol10} obtained a selection rule
for the three-phonon scattering rates stating that only an even number of ZA phonons is attributed to each process. The  four allowed processes involving flexural phonon-modes have been listed by Shen {\it et. al.} \cite{shen14}. Therefore, the scattering rate given by Eq. \ref{tauU} needs to be multiplied by three for the case of MLG. Our calculations suggest that the LA and TA modes contribute maximum to the total $\kappa_L$.

\begin{table}[t]
\centering
\caption{\label{para} Parameters used in the analytical solutions of the Callaway-Klemens method.}
\begin{tabular}{c|c|c|c|c|c|c}
\hline\hline
System & \begin{tabular}{c} $v_{LA}$ \\ (m/s)\end{tabular} & \begin{tabular}{c} $v_{TA}$ \\ (m/s)\end{tabular} & $\gamma_{LA}$ & $\gamma_{TA}$ &\begin{tabular}{c} $\alpha$  $\times$ 10$^{-7}$ \\ (m$^2$/s)\end{tabular} & \begin{tabular}{c} $\beta$ $\times$ 10$^{-20}$\\ (1/m$^2$)\end{tabular} \\
\hline
MLG & 18021.5 & 12968.4 & 1.70 & 0.65 & 5.64  & -7.7  \\
BLG & 18014.3 & 12624.9 & 1.75 & 0.72 & 5.89  & -7.47  \\
\hline\hline
\end{tabular}
\end{table}

\subsection{Lattice thermal conductivity using the Iterative method}

\begin{figure}[!htbp]
\centering \includegraphics[scale=0.32]{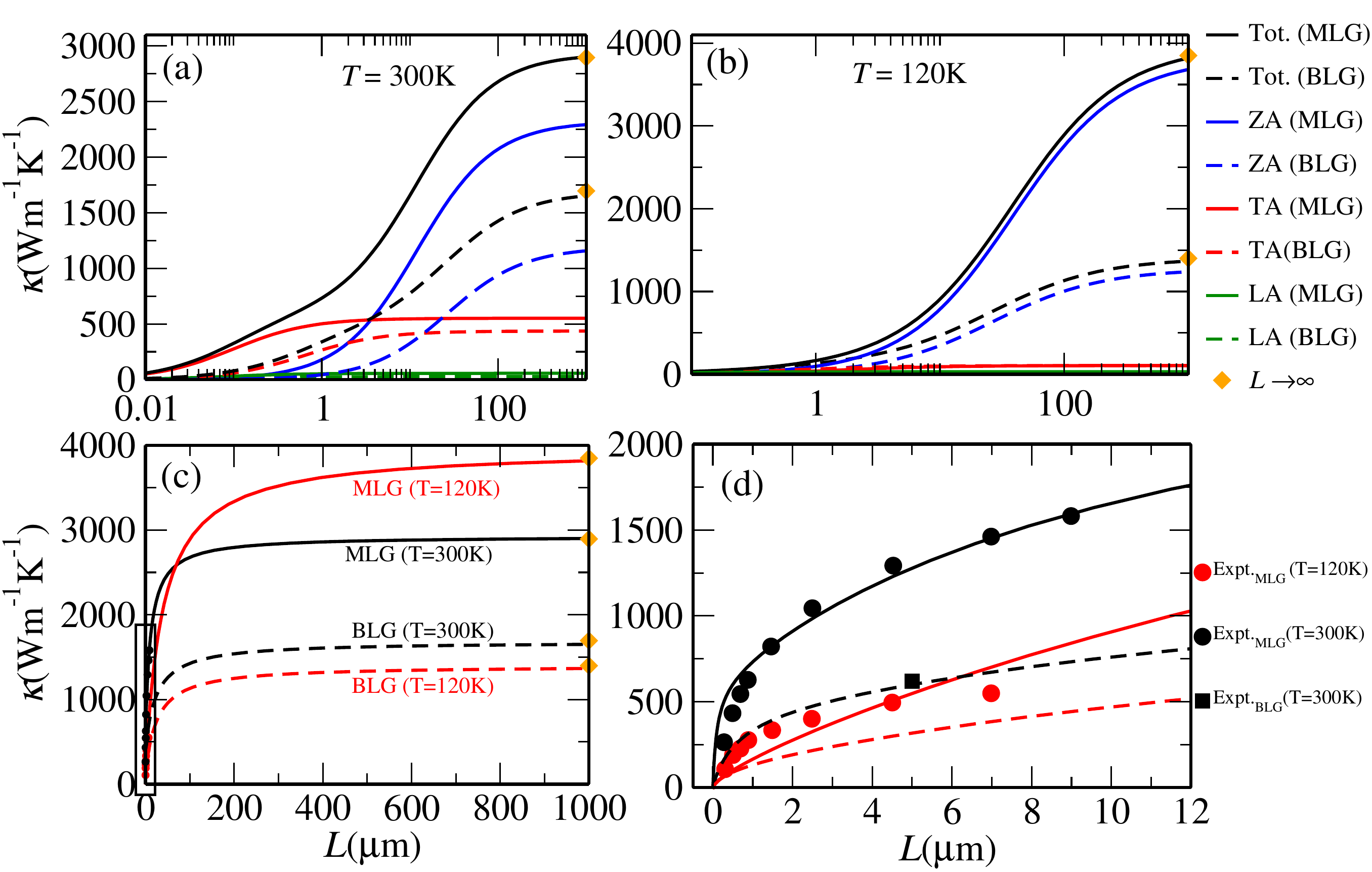}
\caption{\label{ki} The calculated mode-dependent ($\kappa_L$) and total $\kappa_L$ plotted as a function of sample-length $L$ in logarithmic scale(a,b) and linear scale (c,d) at two temperatures, $T=120$K and $T=300$K. Solid(dashed) curves refer to calculation on MLG(BLG). The orange diamond points are the values of $\kappa_L$ \cite{RDSM17} at the thermodynamic limit ($L\rightarrow\infty$). (d) Zoomed box in (c) comparing our calculations with available experimental data \cite{pettes11,xu14}.}
\end{figure}

Figs. \ref{ki}(a) and \ref{ki}(b) show, in logarithmic scale, the length dependence of the contribution from each of the acoustic modes (LA,TA,ZA) to the total lattice thermal conductivity ($\kappa_L$) at two fixed temperatures, $T=300$K and $T=120$K, respectively, for MLG and BLG, calculated using the iterative ShengBTE method \cite{ShengBTE}.
The iterative method clearly shows that the out-of-plane acoustic ZA mode contributes the most to the total lattice thermal conductivity. At room temperature (RT) and at the thermodynamic limit, the contribution for MLG (BLG) are 79\% (70\%), 19\% (26\%) and 2\% (4\%) from the ZA, TA and LA modes, respectively.

We find for BLG, a $\sim$ 9\% drop in $\kappa_L$ in comparison to that of MLG due to the ZA mode.
The major difference between the phonon dispersions between MLG and BLG is the additional out-of-plane optical mode ZO$^{'}$. Due to this additional low-frequency mode, more phase space states are now available for phonon scattering and is one of the reason for the decrease in $\kappa_L$ in BLG.
As evident from Fig. \ref{ki}(b), we find at small sample lengths and at lower temperatures, the mode dependent contributions to $\kappa_L$ are identical for both MLG and BLG indicating that the phonon transport is ballistic and independent to the number of layers.

In Fig. \ref{ki}(c), we plot the total $\kappa_L$ as a function of sample length. Since most of the lattice thermal conductivity measurements were carried out at small sample lengths, in order to compare our calculations to experimental data, we show in Fig. \ref{ki}(d) the zoomed data in the thin rectangular box of Fig. \ref{ki}(c) where experimental measurements are available for the given sample length range.
 
The orange diamond points shown in Figs. \ref{ki}(a), \ref{ki}(b) and \ref{ki}(c) are values of $\kappa_L$ at the thermodynamic limit, reported previously \cite{RDSM17}, at the corresponding temperatures.
The thermodynamic value of $\kappa_L$ for MLG at 120K is higher than that its value at RT while reverse is case for BLG.
This suggests that the temperature dependence of $\kappa_L(T)$ has a peak closer to $T=$120K for MLG, whereas this peak shifts to a higher value for BLG.
Lindsay {\it el. al.} \cite{lindsay14} have shown that the mode dependence of $\kappa_L$ for MLG depends only slightly on strain. The length dependent calculations of $\kappa_L$ using 
 first principles calculation based on DFPT \cite{lindsay14} at RT, referring to each of the acoustic modes, for the unstrained MLG are in very good agreement with our calculations shown in Fig. \ref{ki}(a). 
Our length dependent $\kappa_L$  calculations are in excellent agreement with earlier theoretical calculations \cite{fugallo14} shown in Fig. \ref{ki}(d).

\begin{figure}[!htbp]
\centering \includegraphics[scale=0.32]{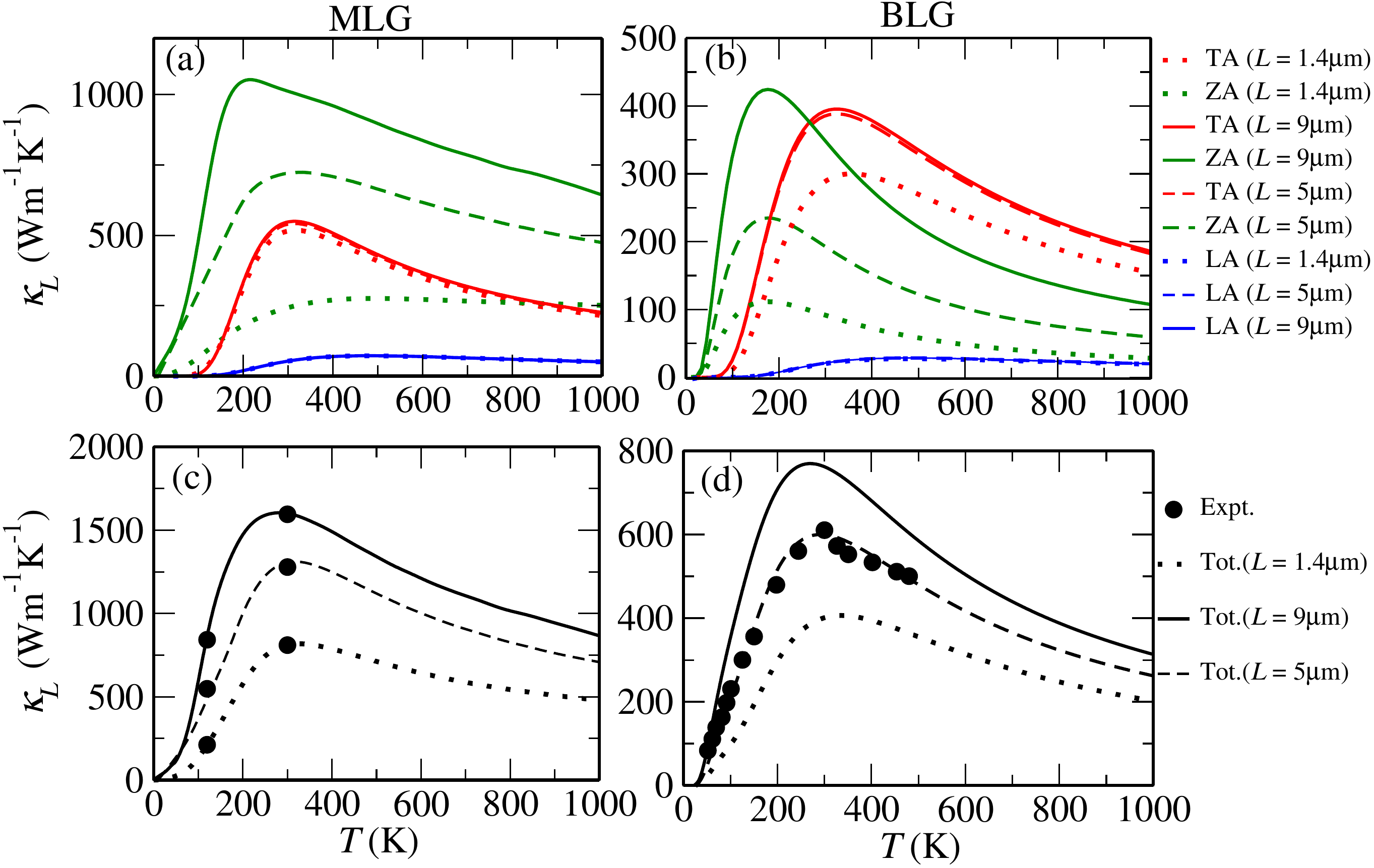}
\caption{\label{kT} The calculated mode-dependent contributions to $\kappa_L$ at three different sample lengths (a,b), and total $\kappa_L$ (c,d) as a function of temperature. Black circular dots are available experimental data \cite{pettes11,xu14}.}
\end{figure}

Fig. \ref{kT} shows the temperature dependence of each of the acoustic modes and the total lattice thermal
conductivity at three constant lengths, $L$=1.4 $\mu$m, 5 $\mu$m and 9 $\mu$m, calculated using the ShengBTE method \cite{ShengBTE}, along with the available experimental data.
The ZA out-of-plane mode is shown to be the most sensitive to length as compared to the in-plane, LA and TA modes. This suggests that the ZA phonons travel ballistically in the sheets while the TA and LA modes travel diffusively.
Measurements of graphene \cite{seol10} on a SiO$_2$ substrate show a reduction in $\kappa_L$ which has been explained with a scattering model where the contributions from the out-of-plane are the most dominant, in line
with our calculations.

Experimental results at the thermodynamic limit ($L\rightarrow\infty$) of $\kappa_L$ at room temperature for graphite show a value of $\sim$ 2000 Wm$^{-1}$K$^{-1}$\, \cite{fugallo14}. Our calculated thermodynamic limit of $\kappa_L$ for BLG is $\sim$ 1700 Wm$^{-1}$K$^{-1}$. This proximity of $\kappa_L$ between BLG and graphite implies that the interlayer interactions are short ranged.

In Figs. \ref{kT}(a) and \ref{kT}(b), it can be seen that at low temperatures, the ZA mode is always larger than the in-plane acoustic modes (LA,TA). This behavior can be understood by considering the phonon density of states (PDOS) which is proportional the number of phonon-modes per frequency interval \cite{lindsay14}.
Using the definition of the 2D density of states, $D_s(\omega)$ $\propto$ $\frac{q}{2\pi}\frac{dq}{d\omega}$, one can measure the contributions from each phonon modes to the total thermal conductivity. Denoting $D_{o}$ and $D_{i}$ as the PDOS for the out-of-plane and in-plane modes, it can be easily shown that, assuming a quadratic ($\omega_i=\alpha q^2$) and linear ($\omega_i=v_i q$) fit to the out-of-plane and in-plane phonon modes, respectively, $\frac{D_{o}}{D_{i}}=\frac{v^2_{i}}{2\alpha\omega_i}$. Where, $v_i$, $\omega_i$ ($i$=LA,TA), are the fitting parameters to the phonon velocity and phonon frequency shown in Table \ref{para} and is plotted in Fig. \ref{phdos}.
Substituting the values from Table \ref{para}, it is evident that at the long wavelength limit ($q \rightarrow 1$), $\frac{D_{o}}{D_{i}}\gg1$.

\subsection{Comparison between Callaway-Klemens and Iterative method}
Though the calculations of the total $\kappa_L$ for MLG using the Callaway-Klemens and the iterative ShengBTE methods exhibit excellent agreement, the results on mode-dependent $\kappa_L$ differ sigficantly.
Our analytical solutions (see Appendix) using the Callaway-Klemens method suggest that, due to the quadratic nature of phonon dispersion of the out-of-plane ZA phonon-mode and the large negative values of Gr\"{u}neisen parameters, contribution from the ZA mode should contribute the least to the total lattice thermal conductivity. On the other hand the LA and TA phonon-modes exhibit smaller Gr\"{u}neisen parameters and linear phonon dispersion, making the phonon group velocities almost constant and large along the boundary of the Brouillon zone.

The iterative ShengBTE method yields, due to the PDOS and symmetry of MLG, the ZA phonon-modes to contribute the most in the total $\kappa_L$ as discussed in the previous subsection. 
Therefore, calculating the thermal conductivity of BLG using the two mentioned methods should end this disparity since the selection rules for the ZA modes is broken for BLG as compared to that in MLG.
It should be noted that, in both methods, the difference between the total $\kappa_L$ in MLG and BLG is due to the contributions of the ZA modes. 
Note, that a multiplicative factor of three in the scattering rate (Eq.\ref{tauU}) in Callaway-Klemens method for MLG was introduced heuristically which is not an artefact of the theory, as discussed in Section III(B).
Absence of the multiplicative factor would imply that the thermal conductivity of MLG would be similar to that of BLG since the magnitude of phonon group velocity and Gr\"{u}neisen parameters are similar for both the materials.
Calculations of $\kappa_L$ by Kong {\it et. al.} \cite{kong2009} have reported that $\kappa^{\rm BLG}_L$ $\sim$ $\kappa^{\rm MLG}_L$, in line with our calculations using the analytical form without the multiplicative factor \cite{RSBN17}. 
However, lattice thermal conductivity experimental data at RT yield $\kappa^{\rm BLG}_L$ $\sim$ 0.68 $\kappa^{\rm MLG}_L$ \cite{hongyang2014}. The iterative ShengBTE method at RT yields $\kappa^{\rm BLG}_L$ $\sim$ 0.60 $\kappa^{\rm MLG}_L$. As seen from Fig.\ref{kfig}, at room temperatures calculations using the Callaway-Klemens method overestimates $\kappa_L$ for BLG. 

Moreover, length dependence calculations of $\kappa_L$ of MLG at 120K using the iterative method are in decent agreement with experimental data at the same temperature, unlike in Callaway-Klemens method where point defects with additional fitting parameters were required to fit the calculations with the experiment (Fig. \ref{kfig}).
Since our calculations using the iterative method are in excellent agreement with the experimental measurements at different lengths and temperatures \cite{hongyang2014,xu14,pettes11} along with other available theoretical calculations based on Tersoff empirical
inter-atomic potentials \cite{lindsay2010,lindsay2011} and first-principles calculations \cite{lindsay14}, the relaxation times calculated by iterative method are accurate as compared to those calculated by the Callaway-Klemens method. 
Calculations of $\kappa_L$ implementing the iterative method on layered hexagonal boron-nitride have shown to be in excellent agreement using the Tersoff potentials \cite{lindsay11,lindsay12} as well as first-principles methods \cite{RSBN17}.
We therefore use only $\kappa_L$  derived from the iterative method for the calculations of the figure of merit ($ZT$) in the next section.

\subsection{Figure of Merit of undoped MLG and BLG}
\begin{figure}[!htbp]
\centering \includegraphics[scale=0.32]{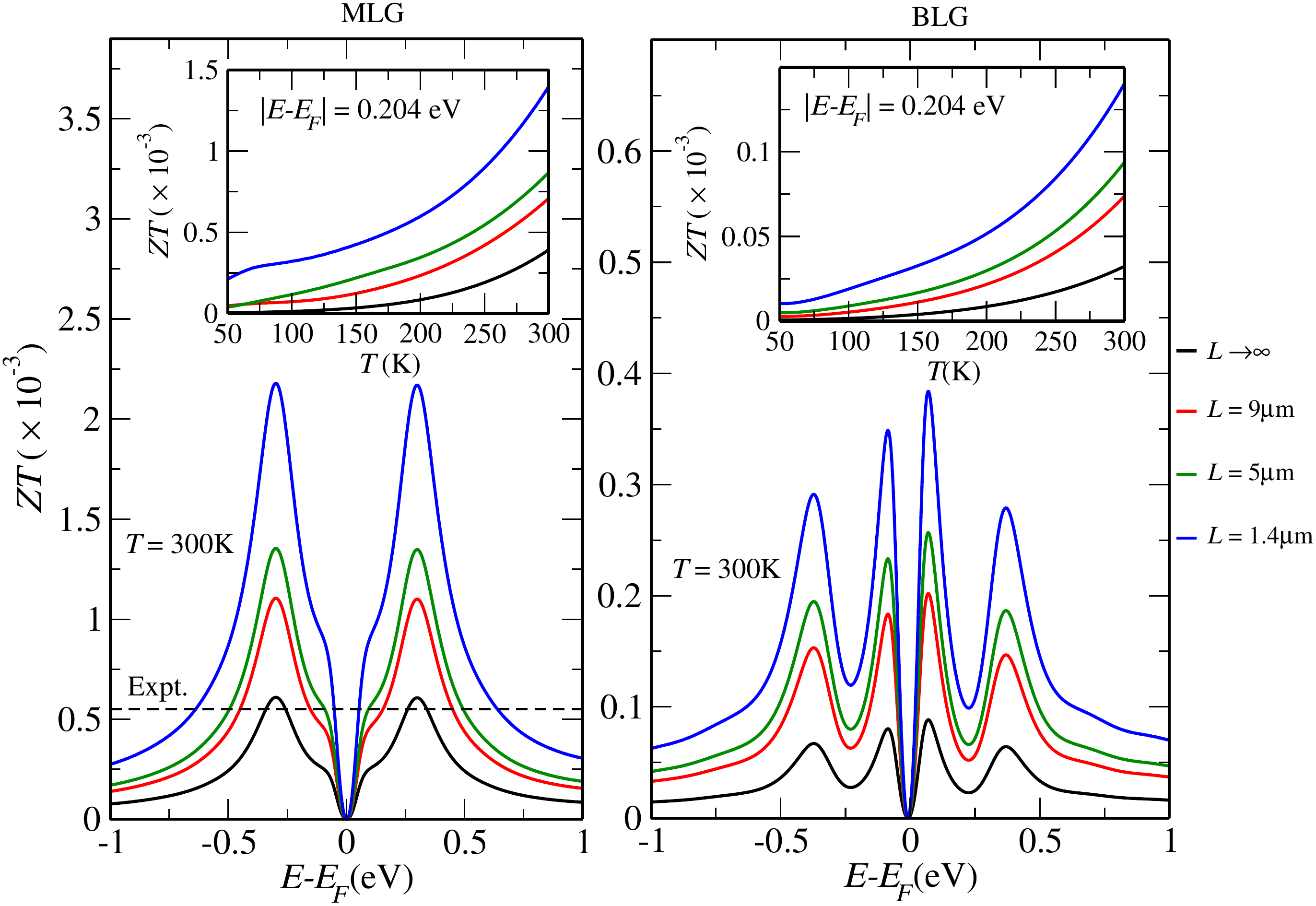}
\caption{\label{ZT}  The calculated Figure of merit ($ZT$) for undoped MLG (left) and BLG (right) at three different lengths, $L$=1.4$\mu$m, $L$=5$\mu$m and $L$=9$\mu$m together with its thermodynamic limit ($L\rightarrow\infty$). 
The black dashed line refers to the experimental data \cite{anno17}. 
Inset: Calculated $ZT$ as a function of temperature for fixed chemical potential.}
\end{figure}

Fig. \ref{ZT} shows the figure of merit ($ZT$) of undoped MLG and BLG at three different lengths together with thermodynamic limit ($L\rightarrow\infty$).
Our calculated $ZT_{max}=0.60 \times 10^{-3}$ at $T=300K$ in thermodynamic limit appears to be in good agreement with recent experimental data available for pristine graphene \cite{anno17}, $ZT$=0.55 $\times 10^{-3}$ (shown as black dashed line).
The electrical Boltzmann transport equations using the RTA yields an electrical relaxation time ($\tau_e$) scaled electrical conductivity ($\frac{\sigma}{\tau_e}$). 
Berger {\it et. al.} \cite{berger06,fang15} have experimentally measured the resistivity to be $\rho=1\ \mu \Omega$cm which results in an electrical conductivity $\approx$ 7.1 $\times 10^7$ $\frac{1}{\Omega{\rm m}}$.
Adopting a Drude model, Tan {\it et. al.} \cite{tan07} have estimated $\tau_e$ as a function of charge density having values in the range 10fs-100ps. 
Using the lower bound for the relaxation time we obtain $\sigma$, as calculated by us recently \cite{RDSM17}, in the same range as seen experimentally \cite{berger06}. Therefore, we use $\tau_e$ = 10 fs in all our calculations for the estimate of the figure of merit for MLG.

It should be noted that in general $\tau_e$ is a function of temperature and the electron momentum and therefore, depends on the direction in the Brillouin zone. Durczewski {\it et al.} \cite{durczewski2000} have devised a formalism to calculate the electron relaxation time and Zahedifar {\it et al.} \cite{zahedifar18} have used this model to calculate the figure of merit of half-Heusler semiconductors. A realistic model of $\tau_e$ is of great importance but difficult to be taken into account. The implementation of $\tau_e$ in the Boltzmann transport equations in the BoltzTrap code \cite{boltztrap1} are treated to be isotropic based on  the formulation of Schulz {\it et al. } \cite{schulz92}. Moreover, the behavior of the electrical resistivity, conductivity, mobility and Seebeck coefficients calculated using an isotropic $\tau_e$ \cite{RDSM17} are in excellent agreement to various experimental measurements \cite{kim,kim2,novoselov05,morozov08}, which indicates that an isotropic $\tau_e$ can be a good approximation for graphene and related materials.

In the inset of Fig. \ref{ZT} we see that the for a fixed chemical potential, in the temperature range 50-300K, $ZT$ is larger for smaller sample lengths.
Our calculated $ZT$ for both, MLG and BLG, are symmetric along the chemical potential.
Due to the linear and parabolic electronic bandstructure of MLG and BLG, respectively; MLG has one peak while BLG has two in their $ZT$ as a function of chemical potential.
Both, MLG and BLG are semi-metals and hence transport would occur only near the Fermi energy because for electrons away from the Fermi energy, there are no available states within a small energy window. 
At the Fermi energy, the $ZT$ is zero because the electronic density of states corresponding to chemical potential at the Fermi energy is zero.

\subsection{Decrement of $\kappa_L$ and Enhancement of $ZT$ in BN-doped MLG}

Defects are commonly considered to be destructive to the properties of a material used in solid states devices.
Nonetheless, defects can occasionally be useful in supplying dopants to control their carrier concentration depending on the carriers either being $n$-type or $p$-type. \cite{araujo12}.
Systems such as graphene have defects introduced in them for technological applications. 
Point defects arise within the planes of graphene mostly in the form of impurity atoms and lattice vacancies.
Foreign impurities such as boron and nitrogen are common $p$-type and $n$-type dopants for graphene. 

Micro-Raman spectroscopy is a method to characterize in-plane defects in graphene-like systems \cite{genc17,araujo12,jorio11}.
The disorder-induced band, also known as the D-band, is a characteristic Raman feature in graphene-like systems.
The D-band has no intensity in the absence of any defects and any given impurity that breaks the translation symmetry of the lattice introduces a D-band intensity in the Raman spectrum.
Along with the D-band, the G-band in Raman spectrum also gives information in understanding defects in graphene-like materials predominantly when the impurity atoms dopes the material to change the bonding strength of the foreign species in the host carbon atom.
Therefore, the ratio of the intensity of the D band to the G-band ($\frac{I_D}{I_G}$)in the Raman spectrum plays an vital role in understanding the defects due to impurity scattering in graphene-like systems.

Study of the disorder due to defects in graphene caused by low energy Ar$^{+}$ ion bombardment  was done by Lucchese {\it et al.}\cite{lucchese10} using Raman scattering. This was carried out by varying the densities of the defects induced with different doses in the ion bombardment. The results of the experiment were modelled by inferring that a single impact of an ion on the graphene sheet would modify the sheet on two length scales \cite{jorio11}.
The model is known as {\it the local activation model}. The two length scales are referred to as $r_A$ and $r_S$ which are the radii of two circular areas measured from the impact point as shown in Fig. \ref{bn-defect}.
The shorter radius, $r_S$, is the structural disorder from the impact point and is know as the structurally-disordered region or the S-region. 
At distance for radii greater $r_S$ but smaller than $r_A$ causes a mixing of Bloch states near the K point and hence enhances the intensity in the D-band in the Raman spectrum. This region is termed as the activated or A-region beyond which the lattice structure is preserved and absent from any defect or impurity \cite{jorio11}.

\begin{figure}[!htbp]
\centering \includegraphics[scale=0.11]{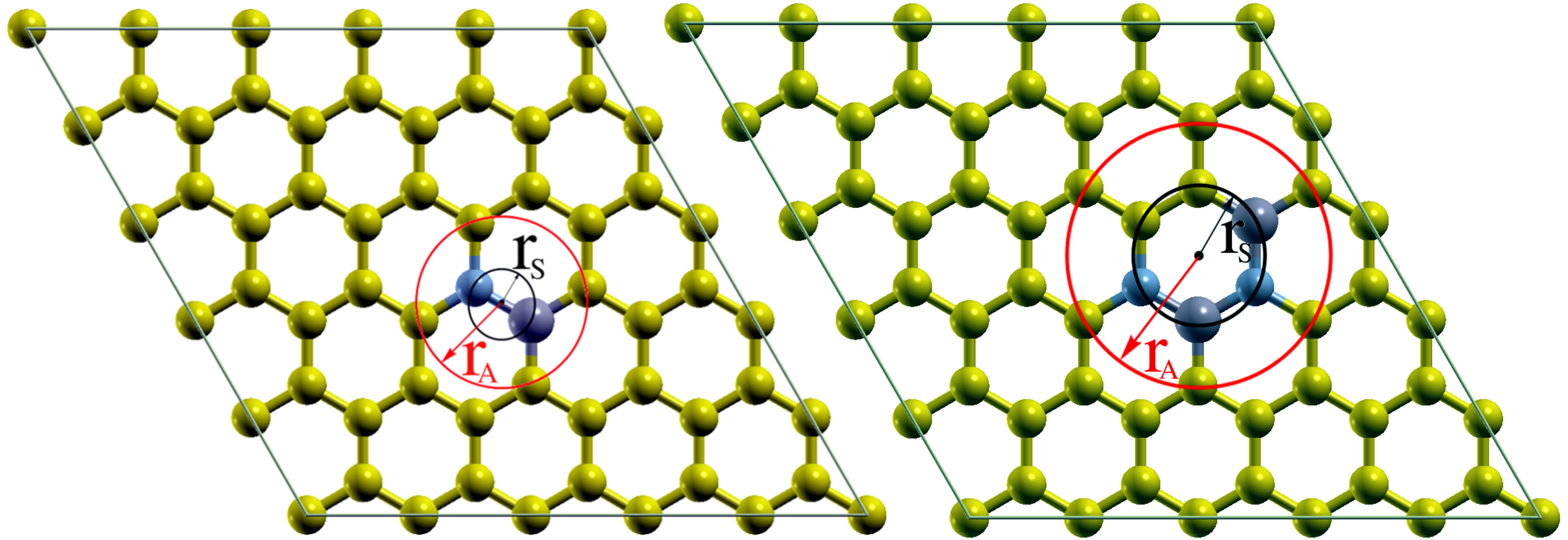}
\caption{\label{bn-defect} Unit cell containing 50 atoms with one BN-dimer (left) and two BN-dimers (right) embedded in graphene used in our calculation. The red and black arrow correspond to the radius of the activated region ($r_A$) and the structurally defective ($r_S$) region, respectively.}
\end{figure}

The local activation model for the $\frac{I_D}{I_G}$ ratio is a function of the average distance between two defects, $L_D$ and is expressed as \cite{jorio11, araujo12,anno17},
\begin{eqnarray}
\frac{I_D}{I_G} &=& C_A\frac{r_A^2-r_S^2}{r_A^2-2r_S^2}\Bigg[e^{\frac{-\pi r_S^2}{L_D^2}}-e^{\frac{-\pi (r_A^2-r_S^2)}{L_D^2}}\Bigg]\nonumber \\ &+&C_S\Bigg[1-e^{\frac{-\pi r_S^2}{L_D^2}}\Bigg]
\end{eqnarray}

where $C_A$ and $C_S$ are adjustable dimensionless parameters. For graphene-like materials the value of $C_A$ and $C_S$ are found to be 4.2 and 0.87 respectively \cite{jorio11,araujo12,anno17} and are the values used in our calculations.
The remaining parameters used in our paper are as follows: The average length between the defects is the same as the length of the unit cell, $L_D = 12.33$\AA. The radii used for the A-region and S-region for one BN-dimer are $r_A = 1.85$\AA\ and $r_S = 0.722$\AA, respectively. Similarly, the radii used for the A-region and S-region for two BN-dimers are $r_A = 2.69$\AA\ and $r_S = 1.44$\AA\, respectively (See Fig.\ref{bn-defect}). 
The resulting $\frac{I_D}{I_G}$ ratio of one and two BN-dimers are calculated to be 0.253 and 0.451, respectively.
The method to calculate the lattice thermal conductivity of BN-doped graphene will be discussed shortly and the results are plotted in the inset of Fig. \ref{dBN-ZT} as a function of the  $\frac{I_D}{I_G}$ ratio.

We have used our previous results \cite{RDSM17} on electrical transport of BN-doped MLG obatined using first-principles DFT based electronic band structure and Boltzmann transport equations for the band electrons for obtaining $\sigma$ and $S$, which are then used to evaluate $ZT$.

For calculating $\kappa_L$ for BN-doped MLG, we have used the iterative ShengBTE method taking BN dimer as point defects in graphene sheets. Calculation of the thermal conductivity of doped MLG was performed by extracting the phonon frequency ($\omega$) dependent phonon relaxation time from the iterative method for MLG, adding the $\omega$-dependent point defects (Eqs. \ref{tpdLATA},\ref{tpdZA}) with calculated parameters, and solving Eqs. \ref{kLATA} and \ref{kZA} with the new calculated phonon relaxation time derived from the Matthiessen’s rule (Eq. \ref{tau}). The required parameters for one and two BN-doping were calculated to be $\Gamma_0^{\rm BN}=7.48\times10^{-4}, \Gamma_0^{\rm 2BN}=1.48\times10^{-3}$, which enter in Eqs. \ref{tpdLATA},\ref{tpdZA}.

Polanco {\it et al.} \cite{polanco18} have calculated the scattering rates due to point defects by various atoms including boron and nitrogen in graphene. The point defect formula used in their paper is very similar to the Eqs. \ref{tpdLATA} and \ref{tpdZA}, with a linear fit (for the LA,TA modes) and a quadratic fit (for the ZA mode) to the phonon dispersion, 
as done by Lindsay {\it et. al.} \cite{lindsay14}. 
\begin{figure}[!htbp]
\centering \includegraphics[scale=0.35]{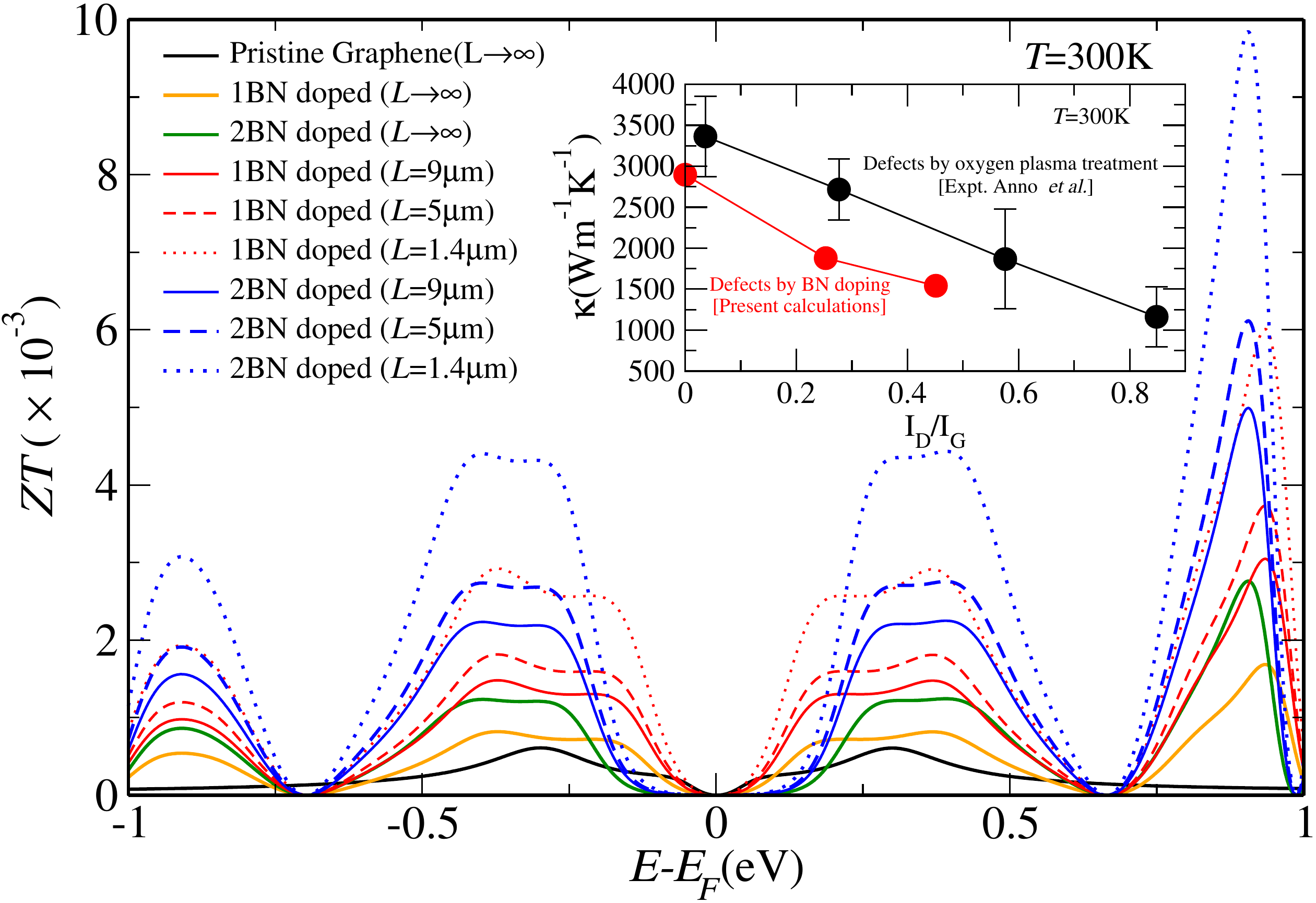}
\caption{\label{dBN-ZT}  The calculated Figure of merit ($ZT$) of one and two BN dimers doped Graphene at 
three different lengths, $L$=1.4$\mu$m, $L$=5$\mu$m and $L$=9$\mu$m along with its thermodynamic limit.
Inset: The lattice thermal conductivity plotted as a function of the ratio of intensity of the D and G band. The black circular points refer to the experimental data of Anno {\it et al.}\cite{anno17}. The red circular points refer to the present calculations.}
\end{figure}

In Fig. \ref{dBN-ZT} we show the figure of Merit for MLG doped with one and two BN dimers at three different sample lengths, $L$=1.4$\mu$m, $L$=5$\mu$m and $L$=9$\mu$m along with its thermodynamic limit. 
The two fold increase in $S$, thereby increasing $ZT$, for MLG upon doping \cite{RDSM17} is attributed to the occurance of a small band gap. Further increase in $ZT$ for smaller sample-lengths is attributable to decrease in $\kappa_L$ as shown in earlier sections.
In the inset of Fig. \ref{dBN-ZT} we plot the lattice thermal conductivity as a function of the  $\frac{I_D}{I_G}$ ratio. We find that the $\kappa_L$ behavior as a function of $\frac{I_D}{I_G}$ ratio is in decent agreement with experiments introducing defects in graphene using oxygen plasma treatment \cite{anno17}.

Our calculations predict that $ZT$ is almost symmetric around the Fermi energy showing an increase with gate voltage for both n(p)-type doping. As one goes to higher values of energy (or gate voltage), there are additional peaks in $ZT$ separated by minima at around 0.7eV above and below the Fermi energy.  The higher values of $ZT$ found at various energy range may lead to increased thermoelectric performance of doped graphene based devices.

Graphene is semi-metallic and gapless which leads to extremely small thermoelectric power factor ($S^2\,\sigma$).
However, a band gap is created at the Fermi level when graphene is doped simultaneously with boron and nitrogen \cite{SMTK2012}, which leads to enhancement in its  thermoelectric power factor. 
Elaborate work have been carried out by various groups \cite{rdsm15,rst17,grossman12} on band gap engineering of boron-nitride doped graphene by varying their constituent concentration. 
Therefore, graphene doped with two BN dimers have negligible $ZT$ around the Fermi level for a larger chemical potential
 range as compared to MLG doped with one BN dimer which in turn has negligible $ZT$ for a larger chemical potential range as compared to pristine MLG.
 This is the product of two BN dimers doped MLG having band gaps greater than one BN dimer doped MLG and that pristine MLG has a no band gap.
 In order to have a better understanding of all the peak seen when $ZT$ is plotted versus the chemical potential, we performed a model calculation described in Appendix B.

\section{Summary}
In summary, using first-principles DFT based electronic and phonon band structure methods together with Boltzmann transport equations for electron and phonon, we have calculated the phonon dispersion and Gr\"{u}neisen parameters for MLG and BLG and find our results to be in good agreement with experimental data (Raman spectroscopy and HREELS). Making a linear and quadratic fit to the in-plane and out-of-plane acoustic phonon dispersion along with constant in-plane and an out-of-plane inverse square wave vector dependent
Gr\"{u}neisen parameters, we find an analytical solution to the mode, length and temperature dependent lattice 
thermal conductivity for the Callaway-Klemens method. The Callaway-Klemens method suggests that the out-of-plane ZA modes contribute the least to the total lattice thermal conductivity due to the large negative Gr\"{u}neisen parameters and vanishing velocities at the long wavelength limit and that the major contribution to $\kappa_L$ are due to the in-plane modes, LA and TA, due to their large velocities and small Gr\"{u}neisen parameters.
The lattice thermal conductivity was also calculated beyond the RTA using an iterative method implemented in the ShengBTE code. The iterative method suggests that, in direct contrast to the Callaway-Klemens method, that the 
ZA modes contribute the most to the total $\kappa_L$ while the in-plane modes contribute the least.

The Callaway-Klemens and iterative method both yield excellent agreement to total $\kappa_L$ for MLG at RT and is the reason as to which mode contributes the most to $\kappa_L$. In order to solve this discrepancy, we calculate the mode, length, and temperature dependent $\kappa_L$ of BLG since the selective rule is broken in the ZA modes for BLG. We find that the Callaway-Klemens method overestimates the thermal conductivity and additional point defects parameters are required to make the theory fit with the experiments. However, using the iterative method, we observe that all our calculations are in excellent agreement with many available experiments without the use of any fitting parameters. We therefore conclude that the thermal conductivity has its major contribution from the ZA mode and is also the most sensitive to the sample length. We also conclude that the mode dependent relaxation time calculated from the Callaway-Klemens method are not accurate and one must go beyond the RTA to solve the relaxation times especially for 2D materials like MLG and BLG.

Along with the electrical transport parameters like electrical conductivity, Seebeck coefficient and hence power factor calculated by us earlier, we have calculated the figure of merit of MLG and BLG. The lattice thermal conductivity used in our calculations were only taken from the iterative method. Our calculation for pristine graphene at the thermodynamic limit are in excellent agreement with available experimental data. We also find an enhancement of the figure of merit when the sample lengths are in order of ~$\mu$m as compared to that of the thermodynamic limit. 
Implementing the activation model, our estimate of the $\frac{I_D}{I_G}$ ratio for graphene doped with one and two BN-dimers treated as point defects are in excellent agreement with an experiment where defects were introduced by oxygen plasma treatment.
Finally, we show that when pristine graphene is doped with one or two boron nitride dimers, the figure of merit is found to be enhanced over a wide range in chemical potential. We have therefore found a new route to enhance the figure of merit of graphene and hence improve graphene based devices over a wide range in gate voltage.

\section{Acknowledgments}
We would like to thank J. Carrete, D.L. Nika and A. Balandin for their helpful correspondance. The calculations were performed in the High Performance Cluster platform of the S.N. Bose National Centre for Basic Sciences (SNBNCBS). RD acknowledges support from SNBNCBS through a Senior Research Fellowship.

\section{Appendix}
\subsection{Simple analytic expression for $\kappa_L$}

In order to have an analytical expression for $\kappa_L$, we make the approximations - 
(i) a linear phonon dispersion for the in-plane acoustic modes, (ii) a quadratic dispersion for the out-of-plane mode, (iii) constant Gr\"{u}neisen parameters corresponding to the in-plane acoustic modes, and (iv) an inverse square wave-vector dependent Gr\"{u}neisen parameters corresponding to the out-of-plane acoustic mode.

For the case of an ideal 2D material {\it i.e.}, a material without any point defects having a specularity parameter of 1, with these approximations, substituting equation \ref{tauU} in \ref{k} it can be shown that the contribution to the total $\kappa_L$ from LA, TA and ZA modes will all have a closed form \cite{nika09, RDSM17}.

We now extend our calculations with point defects and specularity parameters for values of $p$ less than 1.
With these approximations, Eq. \ref{k} for the LA,TA and ZA mode can be easily shown to be,
\begin{eqnarray}
\kappa_{L_{LA,TA}} &=& \frac{1}{C_0} \int\limits_{\omega_{min}}^{\omega_D} \hbar^2 \omega^3 \tau_{tot}
\frac{e^{\frac{\hbar \omega}{k_B T}}}{[e^{\frac{\hbar \omega}{k_B T}}-1]^2} d\omega \label{kLATA}, \\
\kappa_{L_{ZA}} &=& \frac{2}{C_0} \int\limits_{\omega_{min}}^{\omega_D} \hbar^2 \omega^3 \tau_{tot}
\frac{e^{\frac{\hbar \omega}{k_B T}}}{[e^{\frac{\hbar \omega}{k_B T}}-1]^2} d\omega \label{kZA},
\end{eqnarray}
where $C_0$ is given by $C_0 = 4 \pi k_B T^2 (N\delta)$ and the separate relaxation times for the in-plane and
 out-of-plane modes with these approximation become,
\begin{eqnarray}
\tau_{U,s}(\omega) &=& \frac{C_1}{\omega^2}\ \ ;\ \ C_1 = \frac{Mv^2\omega_{D_s}}{\gamma^2 k_B T} \Rightarrow [s = {\rm LA, TA}] \\
 &=& C_2 \omega\ \ ;\ \ C_2 = \frac{4M\omega_{D_s}}{\beta^2 \alpha k_B T} \Rightarrow [s = {\rm ZA}] \\
\tau_{B,s}(\omega) &=& C_3\ \ ;\ \ C_3 = \frac{d}{v}\frac{1+p}{1-p} \Rightarrow [s = {\rm LA,TA}] \\
 &=& \frac{C_4}{\sqrt{\omega}}\ \ ;\ \ C_4 = \frac{d}{2\sqrt{\alpha}} \Rightarrow [s = {\rm ZA}] \\
 \tau_{P,s}(\omega) &=& \frac{C_5}{\omega^3}\ \ ;\ \ C_5 = \frac{4v^2}{S_0\Gamma_0} \Rightarrow [s = {\rm LA,TA}] \label{tpdLATA} \\
 &=& \frac{C_6}{\omega^2}\ \ ;\ \ C_6 = \frac{8\alpha}{S_0\Gamma_0} \Rightarrow [s = {\rm ZA}] \label{tpdZA}
\end{eqnarray}

\subsection{Model calculation of $\sigma$ and $S$ of BN-doped Graphene}

The behavior of the $ZT$ can be apprehended by studying the Seebeck coefficient. For semiconductors with small 
energy band gaps, the Seebeck coefficient, at a constant temperature, can be shown to be $S \propto \frac{d}{dE}[\ln\sigma(E)\big\vert_{E_F}]$ \cite{RDSM17,mott}. Where, $\sigma$ is the electrical conductivity. The electrical conductivity as a function of wave vector $\sigma(k)$ was derived from the wave vector dependent velocity ($v(k)=\frac{d\epsilon}{dk}$), $\sigma(k)\propto v(k)^2$. $\epsilon$ is the energy dispersion derived from the electronic bandstructure. The energy dependent electrical conductivity and velocity are then calculated using, $\sigma(\epsilon) = \sum_{i,k}\sigma(k)\frac{\delta(\epsilon-\epsilon_{i,k})}{d\epsilon}$ and 
$v(\epsilon) = \sum_{i,k}v(k)\frac{\delta(\epsilon-\epsilon_{i,k})}{d\epsilon}$, respectively.
The dummy variable $'i'$ corresponds to the band index. In our model calculation, $i$ runs from $i=1$ to $i=4$, two bands below and above the Fermi energy. This sections aims to understand the behavior of $ZT$ and hence all of the constants in our calculations are set to 1.
\begin{figure}[!htbp]
\includegraphics[scale=0.35]{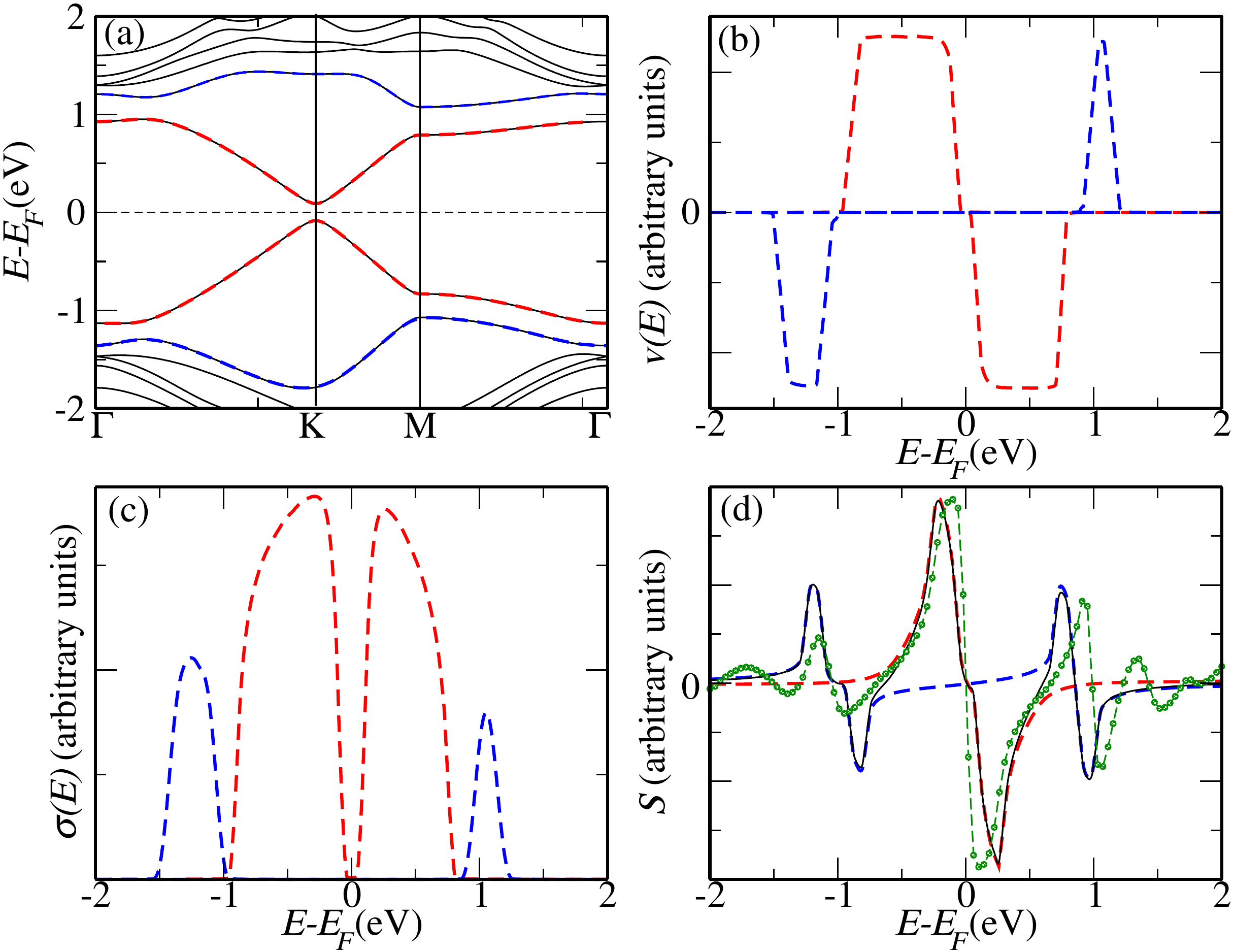}
\caption{\label{dBN-S}(a) Bandstructure of one BN-doped graphene, (b) group velocity of electrons belonging to the two closest bands to the Fermi energy, shown in red and blue, (c) electrical conductivity of these electron system, (d) their Seebeck coefficient. The blue and red curves in (b,c,d) refer to the bands of the same colour as in (a). The green circles in (d) are the first-principles calculations of $S$ taking contributions from all bands \cite{RDSM17}.}
\end{figure} 

Fig. \ref{dBN-S} (a) shows the the bandstructure of one BN dimer doped graphene. The red curves are the two bands closest to the Fermi energy and the blue curves are the next closest.
The energy dependent velocity, electrical conductivity and Seebeck coefficient are plotted in Fig. \ref{dBN-S} (b,c,d) respectively. The colour conventions for these curves correspond to the colour of the bands in Fig.\ref{dBN-S} (a). 
It is evident from our calculations that the zeros in the Figure of Merit are due to the vanishing electron velocities and hence electrical conductivities. Our results using this model calculation show that the 
features at $|E-E_F| \sim $ 1 eV, which are absent in pristine graphene, are due to the bands which are second to the closest bands to the Fermi energy. The form of the Seebeck coefficient shown in Fig. \ref{dBN-S} (d) is in
decent agreement to the form calculated using the Boltzmann equations implemented in the BOLTZTRAP code, shown in green circles\cite{RDSM17,boltztrap1}.

\bibliographystyle{aipnum4-1}

\begin{thebibliography}{76}%
\makeatletter
\providecommand \@ifxundefined [1]{%
 \@ifx{#1\undefined}
}%
\providecommand \@ifnum [1]{%
 \ifnum #1\expandafter \@firstoftwo
 \else \expandafter \@secondoftwo
 \fi
}%
\providecommand \@ifx [1]{%
 \ifx #1\expandafter \@firstoftwo
 \else \expandafter \@secondoftwo
 \fi
}%
\providecommand \natexlab [1]{#1}%
\providecommand \enquote  [1]{``#1''}%
\providecommand \bibnamefont  [1]{#1}%
\providecommand \bibfnamefont [1]{#1}%
\providecommand \citenamefont [1]{#1}%
\providecommand \href@noop [0]{\@secondoftwo}%
\providecommand \href [0]{\begingroup \@sanitize@url \@href}%
\providecommand \@href[1]{\@@startlink{#1}\@@href}%
\providecommand \@@href[1]{\endgroup#1\@@endlink}%
\providecommand \@sanitize@url [0]{\catcode `\\12\catcode `\$12\catcode
  `\&12\catcode `\#12\catcode `\^12\catcode `\_12\catcode `\%12\relax}%
\providecommand \@@startlink[1]{}%
\providecommand \@@endlink[0]{}%
\providecommand \url  [0]{\begingroup\@sanitize@url \@url }%
\providecommand \@url [1]{\endgroup\@href {#1}{\urlprefix }}%
\providecommand \urlprefix  [0]{URL }%
\providecommand \Eprint [0]{\href }%
\providecommand \doibase [0]{http://dx.doi.org/}%
\providecommand \selectlanguage [0]{\@gobble}%
\providecommand \bibinfo  [0]{\@secondoftwo}%
\providecommand \bibfield  [0]{\@secondoftwo}%
\providecommand \translation [1]{[#1]}%
\providecommand \BibitemOpen [0]{}%
\providecommand \bibitemStop [0]{}%
\providecommand \bibitemNoStop [0]{.\EOS\space}%
\providecommand \EOS [0]{\spacefactor3000\relax}%
\providecommand \BibitemShut  [1]{\csname bibitem#1\endcsname}%
\let\auto@bib@innerbib\@empty
\bibitem [{\citenamefont {Geim}\ and\ \citenamefont
  {Novoselov}(2007)}]{geim07}%
  \BibitemOpen
  \bibfield  {author} {\bibinfo {author} {\bibfnamefont {A.~K.}\ \bibnamefont
  {Geim}}\ and\ \bibinfo {author} {\bibfnamefont {K.~S.}\ \bibnamefont
  {Novoselov}},\ }\href@noop {} {\bibfield  {journal} {\bibinfo  {journal}
  {Nature materials}\ }\textbf {\bibinfo {volume} {6}},\ \bibinfo {pages} {183}
  (\bibinfo {year} {2007})}\BibitemShut {NoStop}%
\bibitem [{\citenamefont {Bolotin}\ \emph
  {et~al.}(2008{\natexlab{a}})\citenamefont {Bolotin}, \citenamefont {Sikes},
  \citenamefont {Hone}, \citenamefont {Stormer},\ and\ \citenamefont
  {Kim}}]{kim08}%
  \BibitemOpen
  \bibfield  {author} {\bibinfo {author} {\bibfnamefont {K.~I.}\ \bibnamefont
  {Bolotin}}, \bibinfo {author} {\bibfnamefont {K.~J.}\ \bibnamefont {Sikes}},
  \bibinfo {author} {\bibfnamefont {J.}~\bibnamefont {Hone}}, \bibinfo {author}
  {\bibfnamefont {H.~L.}\ \bibnamefont {Stormer}}, \ and\ \bibinfo {author}
  {\bibfnamefont {P.}~\bibnamefont {Kim}},\ }\href@noop {} {\bibfield
  {journal} {\bibinfo  {journal} {Physical Review Letters}\ }\textbf {\bibinfo
  {volume} {101}},\ \bibinfo {pages} {096802} (\bibinfo {year}
  {2008}{\natexlab{a}})}\BibitemShut {NoStop}%
\bibitem [{\citenamefont {Zuev}, \citenamefont {Chang},\ and\ \citenamefont
  {Kim}(2009{\natexlab{a}})}]{kim09}%
  \BibitemOpen
  \bibfield  {author} {\bibinfo {author} {\bibfnamefont {Y.}~\bibnamefont
  {Zuev}}, \bibinfo {author} {\bibfnamefont {W.}~\bibnamefont {Chang}}, \ and\
  \bibinfo {author} {\bibfnamefont {P.}~\bibnamefont {Kim}},\ }\href@noop {}
  {\bibfield  {journal} {\bibinfo  {journal} {Physical Review Letters}\
  }\textbf {\bibinfo {volume} {102}},\ \bibinfo {pages} {096807} (\bibinfo
  {year} {2009}{\natexlab{a}})}\BibitemShut {NoStop}%
\bibitem [{\citenamefont {Novoselov}\ \emph {et~al.}(2005)\citenamefont
  {Novoselov}, \citenamefont {Geim}, \citenamefont {Morozov}, \citenamefont
  {Jiang}, \citenamefont {Katsnelson}, \citenamefont {Grigorieva},
  \citenamefont {Dubonos},\ and\ \citenamefont {Firsov}}]{novoselov05}%
  \BibitemOpen
  \bibfield  {author} {\bibinfo {author} {\bibfnamefont {K.~S.}\ \bibnamefont
  {Novoselov}}, \bibinfo {author} {\bibfnamefont {A.~K.}\ \bibnamefont {Geim}},
  \bibinfo {author} {\bibfnamefont {S.~V.}\ \bibnamefont {Morozov}}, \bibinfo
  {author} {\bibfnamefont {D.}~\bibnamefont {Jiang}}, \bibinfo {author}
  {\bibfnamefont {M.~I.}\ \bibnamefont {Katsnelson}}, \bibinfo {author}
  {\bibfnamefont {I.~V.}\ \bibnamefont {Grigorieva}}, \bibinfo {author}
  {\bibfnamefont {S.~V.}\ \bibnamefont {Dubonos}}, \ and\ \bibinfo {author}
  {\bibfnamefont {A.~A.}\ \bibnamefont {Firsov}},\ }\href@noop {} {\bibfield
  {journal} {\bibinfo  {journal} {Nature}\ }\textbf {\bibinfo {volume} {438}},\
  \bibinfo {pages} {197} (\bibinfo {year} {2005})}\BibitemShut {NoStop}%
\bibitem [{\citenamefont {Zhang}\ \emph {et~al.}(2005)\citenamefont {Zhang},
  \citenamefont {Tan}, \citenamefont {Stormer},\ and\ \citenamefont
  {Kim}}]{zhang05}%
  \BibitemOpen
  \bibfield  {author} {\bibinfo {author} {\bibfnamefont {Y.}~\bibnamefont
  {Zhang}}, \bibinfo {author} {\bibfnamefont {Y.~W.}\ \bibnamefont {Tan}},
  \bibinfo {author} {\bibfnamefont {H.~L.}\ \bibnamefont {Stormer}}, \ and\
  \bibinfo {author} {\bibfnamefont {P.}~\bibnamefont {Kim}},\ }\href@noop {}
  {\bibfield  {journal} {\bibinfo  {journal} {Nature}\ }\textbf {\bibinfo
  {volume} {438}},\ \bibinfo {pages} {201} (\bibinfo {year}
  {2005})}\BibitemShut {NoStop}%
\bibitem [{\citenamefont {Schedin}\ \emph {et~al.}(2007)\citenamefont
  {Schedin}, \citenamefont {Geim}, \citenamefont {Morozov}, \citenamefont
  {Hill}, \citenamefont {Blake}, \citenamefont {Katsnelson},\ and\
  \citenamefont {Novoselov}}]{schedin07}%
  \BibitemOpen
  \bibfield  {author} {\bibinfo {author} {\bibfnamefont {F.}~\bibnamefont
  {Schedin}}, \bibinfo {author} {\bibfnamefont {A.~K.}\ \bibnamefont {Geim}},
  \bibinfo {author} {\bibfnamefont {S.~V.}\ \bibnamefont {Morozov}}, \bibinfo
  {author} {\bibfnamefont {E.~W.}\ \bibnamefont {Hill}}, \bibinfo {author}
  {\bibfnamefont {P.}~\bibnamefont {Blake}}, \bibinfo {author} {\bibfnamefont
  {M.~I.}\ \bibnamefont {Katsnelson}}, \ and\ \bibinfo {author} {\bibfnamefont
  {K.~S.}\ \bibnamefont {Novoselov}},\ }\href@noop {} {\bibfield  {journal}
  {\bibinfo  {journal} {Nature Mater.}\ }\textbf {\bibinfo {volume} {6}},\
  \bibinfo {pages} {652} (\bibinfo {year} {2007})}\BibitemShut {NoStop}%
\bibitem [{\citenamefont {Tan}\ \emph {et~al.}(2007)\citenamefont {Tan},
  \citenamefont {Zhang}, \citenamefont {Bolotin}, \citenamefont {Zhao},
  \citenamefont {Adam}, \citenamefont {Hwang}, \citenamefont {Sarma},
  \citenamefont {Stormer},\ and\ \citenamefont {Kim}}]{tan07}%
  \BibitemOpen
  \bibfield  {author} {\bibinfo {author} {\bibfnamefont {Y.~W.}\ \bibnamefont
  {Tan}}, \bibinfo {author} {\bibfnamefont {Y.}~\bibnamefont {Zhang}}, \bibinfo
  {author} {\bibfnamefont {K.}~\bibnamefont {Bolotin}}, \bibinfo {author}
  {\bibfnamefont {Y.}~\bibnamefont {Zhao}}, \bibinfo {author} {\bibfnamefont
  {S.}~\bibnamefont {Adam}}, \bibinfo {author} {\bibfnamefont {E.~H.}\
  \bibnamefont {Hwang}}, \bibinfo {author} {\bibfnamefont {S.~D.}\ \bibnamefont
  {Sarma}}, \bibinfo {author} {\bibfnamefont {H.~L.}\ \bibnamefont {Stormer}},
  \ and\ \bibinfo {author} {\bibfnamefont {P.}~\bibnamefont {Kim}},\
  }\href@noop {} {\bibfield  {journal} {\bibinfo  {journal} {Phys. Rev. Lett.}\
  }\textbf {\bibinfo {volume} {99}},\ \bibinfo {pages} {246803} (\bibinfo
  {year} {2007})}\BibitemShut {NoStop}%
\bibitem [{\citenamefont {Chen}\ \emph {et~al.}(2008)\citenamefont {Chen},
  \citenamefont {Jang}, \citenamefont {Adam}, \citenamefont {Fuhrer},
  \citenamefont {Williams},\ and\ \citenamefont {Ishigami}}]{chen08}%
  \BibitemOpen
  \bibfield  {author} {\bibinfo {author} {\bibfnamefont {J.~H.}\ \bibnamefont
  {Chen}}, \bibinfo {author} {\bibfnamefont {C.}~\bibnamefont {Jang}}, \bibinfo
  {author} {\bibfnamefont {S.}~\bibnamefont {Adam}}, \bibinfo {author}
  {\bibfnamefont {M.~S.}\ \bibnamefont {Fuhrer}}, \bibinfo {author}
  {\bibfnamefont {E.~D.}\ \bibnamefont {Williams}}, \ and\ \bibinfo {author}
  {\bibfnamefont {M.}~\bibnamefont {Ishigami}},\ }\href@noop {} {\bibfield
  {journal} {\bibinfo  {journal} {Nature Physics}\ }\textbf {\bibinfo {volume}
  {4}},\ \bibinfo {pages} {377} (\bibinfo {year} {2008})}\BibitemShut {NoStop}%
\bibitem [{\citenamefont {Min}\ \emph {et~al.}(2007)\citenamefont {Min},
  \citenamefont {Sahu}, \citenamefont {Banerjee},\ and\ \citenamefont
  {MacDonald}}]{hongki07}%
  \BibitemOpen
  \bibfield  {author} {\bibinfo {author} {\bibfnamefont {H.}~\bibnamefont
  {Min}}, \bibinfo {author} {\bibfnamefont {B.}~\bibnamefont {Sahu}}, \bibinfo
  {author} {\bibfnamefont {S.~K.}\ \bibnamefont {Banerjee}}, \ and\ \bibinfo
  {author} {\bibfnamefont {A.~H.}\ \bibnamefont {MacDonald}},\ }\href@noop {}
  {\bibfield  {journal} {\bibinfo  {journal} {Phys. Rev. B}\ }\textbf {\bibinfo
  {volume} {75}},\ \bibinfo {pages} {155115} (\bibinfo {year}
  {2007})}\BibitemShut {NoStop}%
\bibitem [{\citenamefont {Repp}\ \emph {et~al.}(2018)\citenamefont {Repp},
  \citenamefont {Harputlu}, \citenamefont {Gurgen}, \citenamefont {Castellano},
  \citenamefont {Kremer}, \citenamefont {Pompe}, \citenamefont {W\"orner},
  \citenamefont {Hoffmann}, \citenamefont {Thomann}, \citenamefont {Emen},
  \citenamefont {Weber}, \citenamefont {Ocakoglubf},\ and\ \citenamefont
  {Erdem}}]{repp18}%
  \BibitemOpen
  \bibfield  {author} {\bibinfo {author} {\bibfnamefont {S.}~\bibnamefont
  {Repp}}, \bibinfo {author} {\bibfnamefont {E.}~\bibnamefont {Harputlu}},
  \bibinfo {author} {\bibfnamefont {S.}~\bibnamefont {Gurgen}}, \bibinfo
  {author} {\bibfnamefont {M.}~\bibnamefont {Castellano}}, \bibinfo {author}
  {\bibfnamefont {N.}~\bibnamefont {Kremer}}, \bibinfo {author} {\bibfnamefont
  {N.}~\bibnamefont {Pompe}}, \bibinfo {author} {\bibfnamefont
  {J.}~\bibnamefont {W\"orner}}, \bibinfo {author} {\bibfnamefont
  {A.}~\bibnamefont {Hoffmann}}, \bibinfo {author} {\bibfnamefont
  {R.}~\bibnamefont {Thomann}}, \bibinfo {author} {\bibfnamefont {F.~M.}\
  \bibnamefont {Emen}}, \bibinfo {author} {\bibfnamefont {S.}~\bibnamefont
  {Weber}}, \bibinfo {author} {\bibfnamefont {K.}~\bibnamefont {Ocakoglubf}}, \
  and\ \bibinfo {author} {\bibfnamefont {E.}~\bibnamefont {Erdem}},\
  }\href@noop {} {\bibfield  {journal} {\bibinfo  {journal} {Nanoscale}\
  }\textbf {\bibinfo {volume} {10}},\ \bibinfo {pages} {1877} (\bibinfo {year}
  {2018})}\BibitemShut {NoStop}%
\bibitem [{\citenamefont {Pham}\ \emph {et~al.}(2016)\citenamefont {Pham},
  \citenamefont {Repp}, \citenamefont {Thomann}, \citenamefont {Krueger},
  \citenamefont {Weber},\ and\ \citenamefont {Erdem}}]{pham16}%
  \BibitemOpen
  \bibfield  {author} {\bibinfo {author} {\bibfnamefont {C.~V.}\ \bibnamefont
  {Pham}}, \bibinfo {author} {\bibfnamefont {S.}~\bibnamefont {Repp}}, \bibinfo
  {author} {\bibfnamefont {R.}~\bibnamefont {Thomann}}, \bibinfo {author}
  {\bibfnamefont {M.}~\bibnamefont {Krueger}}, \bibinfo {author} {\bibfnamefont
  {S.}~\bibnamefont {Weber}}, \ and\ \bibinfo {author} {\bibfnamefont
  {E.}~\bibnamefont {Erdem}},\ }\href@noop {} {\bibfield  {journal} {\bibinfo
  {journal} {Nanoscale}\ }\textbf {\bibinfo {volume} {8}},\ \bibinfo {pages}
  {9682} (\bibinfo {year} {2016})}\BibitemShut {NoStop}%
\bibitem [{\citenamefont {Bonaccorso}\ \emph {et~al.}(2015)\citenamefont
  {Bonaccorso}, \citenamefont {Colombo}, \citenamefont {Yu}, \citenamefont
  {Stoller}, \citenamefont {Tozzini}, \citenamefont {Ferrari}, \citenamefont
  {Ruoff},\ and\ \citenamefont {Pellegrini}}]{Bonaccorso15}%
  \BibitemOpen
  \bibfield  {author} {\bibinfo {author} {\bibfnamefont {F.}~\bibnamefont
  {Bonaccorso}}, \bibinfo {author} {\bibfnamefont {L.}~\bibnamefont {Colombo}},
  \bibinfo {author} {\bibfnamefont {G.}~\bibnamefont {Yu}}, \bibinfo {author}
  {\bibfnamefont {M.}~\bibnamefont {Stoller}}, \bibinfo {author} {\bibfnamefont
  {V.}~\bibnamefont {Tozzini}}, \bibinfo {author} {\bibfnamefont {A.~C.}\
  \bibnamefont {Ferrari}}, \bibinfo {author} {\bibfnamefont {R.~S.}\
  \bibnamefont {Ruoff}}, \ and\ \bibinfo {author} {\bibfnamefont
  {V.}~\bibnamefont {Pellegrini}},\ }\href@noop {} {\bibfield  {journal}
  {\bibinfo  {journal} {Science}\ }\textbf {\bibinfo {volume} {347}},\ \bibinfo
  {pages} {1246501} (\bibinfo {year} {2015})}\BibitemShut {NoStop}%
\bibitem [{\citenamefont {Lindsay}\ \emph {et~al.}(2014)\citenamefont
  {Lindsay}, \citenamefont {Li}, \citenamefont {Carrete}, \citenamefont
  {Mingo}, \citenamefont {Broido},\ and\ \citenamefont {Reinecke}}]{lindsay14}%
  \BibitemOpen
  \bibfield  {author} {\bibinfo {author} {\bibfnamefont {L.}~\bibnamefont
  {Lindsay}}, \bibinfo {author} {\bibfnamefont {W.}~\bibnamefont {Li}},
  \bibinfo {author} {\bibfnamefont {J.}~\bibnamefont {Carrete}}, \bibinfo
  {author} {\bibfnamefont {N.}~\bibnamefont {Mingo}}, \bibinfo {author}
  {\bibfnamefont {D.~A.}\ \bibnamefont {Broido}}, \ and\ \bibinfo {author}
  {\bibfnamefont {T.~L.}\ \bibnamefont {Reinecke}},\ }\href@noop {} {\bibfield
  {journal} {\bibinfo  {journal} {Phys. Rev. B}\ }\textbf {\bibinfo {volume}
  {89}},\ \bibinfo {pages} {155426} (\bibinfo {year} {2014})}\BibitemShut
  {NoStop}%
\bibitem [{\citenamefont {Saito}, \citenamefont {Mizuno},\ and\ \citenamefont
  {Dresselhaus}(2018)}]{saito18}%
  \BibitemOpen
  \bibfield  {author} {\bibinfo {author} {\bibfnamefont {R.}~\bibnamefont
  {Saito}}, \bibinfo {author} {\bibfnamefont {M.}~\bibnamefont {Mizuno}}, \
  and\ \bibinfo {author} {\bibfnamefont {M.~S.}\ \bibnamefont {Dresselhaus}},\
  }\href@noop {} {\bibfield  {journal} {\bibinfo  {journal} {Phys. Rev. Appl.}\
  }\textbf {\bibinfo {volume} {9}},\ \bibinfo {pages} {024017} (\bibinfo {year}
  {2018})}\BibitemShut {NoStop}%
\bibitem [{\citenamefont {D'Souza}\ and\ \citenamefont
  {Mukherjee}(2017{\natexlab{a}})}]{RDSM17}%
  \BibitemOpen
  \bibfield  {author} {\bibinfo {author} {\bibfnamefont {R.}~\bibnamefont
  {D'Souza}}\ and\ \bibinfo {author} {\bibfnamefont {S.}~\bibnamefont
  {Mukherjee}},\ }\href@noop {} {\bibfield  {journal} {\bibinfo  {journal}
  {Phys. Rev. B}\ }\textbf {\bibinfo {volume} {95}},\ \bibinfo {pages} {085435}
  (\bibinfo {year} {2017}{\natexlab{a}})}\BibitemShut {NoStop}%
\bibitem [{\citenamefont {Kuang}\ \emph {et~al.}(2016)\citenamefont {Kuang},
  \citenamefont {Lindsay}, \citenamefont {Shi}, \citenamefont {Wang},\ and\
  \citenamefont {Huang}}]{kuang16}%
  \BibitemOpen
  \bibfield  {author} {\bibinfo {author} {\bibfnamefont {Y.}~\bibnamefont
  {Kuang}}, \bibinfo {author} {\bibfnamefont {L.}~\bibnamefont {Lindsay}},
  \bibinfo {author} {\bibfnamefont {S.}~\bibnamefont {Shi}}, \bibinfo {author}
  {\bibfnamefont {X.}~\bibnamefont {Wang}}, \ and\ \bibinfo {author}
  {\bibfnamefont {B.}~\bibnamefont {Huang}},\ }\href@noop {} {\bibfield
  {journal} {\bibinfo  {journal} {International Journal of Heat and Mass
  Transfer}\ }\textbf {\bibinfo {volume} {101}},\ \bibinfo {pages} {772}
  (\bibinfo {year} {2016})}\BibitemShut {NoStop}%
\bibitem [{\citenamefont {Lindsay}\ and\ \citenamefont
  {Broido}(2011)}]{lindsay11}%
  \BibitemOpen
  \bibfield  {author} {\bibinfo {author} {\bibfnamefont {L.}~\bibnamefont
  {Lindsay}}\ and\ \bibinfo {author} {\bibfnamefont {D.~A.}\ \bibnamefont
  {Broido}},\ }\href@noop {} {\bibfield  {journal} {\bibinfo  {journal} {Phys.
  Rev. B}\ }\textbf {\bibinfo {volume} {84}},\ \bibinfo {pages} {155421}
  (\bibinfo {year} {2011})}\BibitemShut {NoStop}%
\bibitem [{\citenamefont {Lindsay}, \citenamefont {Broido},\ and\ \citenamefont
  {Mingo}(2011)}]{lindsay2011}%
  \BibitemOpen
  \bibfield  {author} {\bibinfo {author} {\bibfnamefont {L.}~\bibnamefont
  {Lindsay}}, \bibinfo {author} {\bibfnamefont {D.~A.}\ \bibnamefont {Broido}},
  \ and\ \bibinfo {author} {\bibfnamefont {N.}~\bibnamefont {Mingo}},\
  }\href@noop {} {\bibfield  {journal} {\bibinfo  {journal} {Phys. Rev. B}\
  }\textbf {\bibinfo {volume} {83}},\ \bibinfo {pages} {235428} (\bibinfo
  {year} {2011})}\BibitemShut {NoStop}%
\bibitem [{\citenamefont {Seol}\ \emph {et~al.}(2010)\citenamefont {Seol},
  \citenamefont {Jo}, \citenamefont {Moore}, \citenamefont {Lindsay},
  \citenamefont {Aitken}, \citenamefont {Pettes}, \citenamefont {Li},
  \citenamefont {Yao}, \citenamefont {Huang}, \citenamefont {Broido},
  \citenamefont {Mingo}, \citenamefont {Ruoff},\ and\ \citenamefont
  {Shi}}]{seol10}%
  \BibitemOpen
  \bibfield  {author} {\bibinfo {author} {\bibfnamefont {J.~H.}\ \bibnamefont
  {Seol}}, \bibinfo {author} {\bibfnamefont {I.}~\bibnamefont {Jo}}, \bibinfo
  {author} {\bibfnamefont {A.~L.}\ \bibnamefont {Moore}}, \bibinfo {author}
  {\bibfnamefont {L.}~\bibnamefont {Lindsay}}, \bibinfo {author} {\bibfnamefont
  {Z.~H.}\ \bibnamefont {Aitken}}, \bibinfo {author} {\bibfnamefont {M.~T.}\
  \bibnamefont {Pettes}}, \bibinfo {author} {\bibfnamefont {X.}~\bibnamefont
  {Li}}, \bibinfo {author} {\bibfnamefont {Z.}~\bibnamefont {Yao}}, \bibinfo
  {author} {\bibfnamefont {R.}~\bibnamefont {Huang}}, \bibinfo {author}
  {\bibfnamefont {D.}~\bibnamefont {Broido}}, \bibinfo {author} {\bibfnamefont
  {N.}~\bibnamefont {Mingo}}, \bibinfo {author} {\bibfnamefont {R.~S.}\
  \bibnamefont {Ruoff}}, \ and\ \bibinfo {author} {\bibfnamefont
  {L.}~\bibnamefont {Shi}},\ }\href@noop {} {\bibfield  {journal} {\bibinfo
  {journal} {Science}\ }\textbf {\bibinfo {volume} {328}},\ \bibinfo {pages}
  {213} (\bibinfo {year} {2010})}\BibitemShut {NoStop}%
\bibitem [{\citenamefont {Lindsay}, \citenamefont {Broido},\ and\ \citenamefont
  {Mingo}(2010)}]{lindsay2010}%
  \BibitemOpen
  \bibfield  {author} {\bibinfo {author} {\bibfnamefont {L.}~\bibnamefont
  {Lindsay}}, \bibinfo {author} {\bibfnamefont {D.~A.}\ \bibnamefont {Broido}},
  \ and\ \bibinfo {author} {\bibfnamefont {N.}~\bibnamefont {Mingo}},\
  }\href@noop {} {\bibfield  {journal} {\bibinfo  {journal} {Phys. Rev. B}\
  }\textbf {\bibinfo {volume} {82}},\ \bibinfo {pages} {115427} (\bibinfo
  {year} {2010})}\BibitemShut {NoStop}%
\bibitem [{\citenamefont {Shen}\ \emph {et~al.}(2014)\citenamefont {Shen},
  \citenamefont {Xie}, \citenamefont {Wei}, \citenamefont {Zhang},
  \citenamefont {Tang}, \citenamefont {Zhong}, \citenamefont {Zhang},\ and\
  \citenamefont {Zhang}}]{shen14}%
  \BibitemOpen
  \bibfield  {author} {\bibinfo {author} {\bibfnamefont {Y.}~\bibnamefont
  {Shen}}, \bibinfo {author} {\bibfnamefont {G.}~\bibnamefont {Xie}}, \bibinfo
  {author} {\bibfnamefont {X.}~\bibnamefont {Wei}}, \bibinfo {author}
  {\bibfnamefont {K.}~\bibnamefont {Zhang}}, \bibinfo {author} {\bibfnamefont
  {M.}~\bibnamefont {Tang}}, \bibinfo {author} {\bibfnamefont {J.}~\bibnamefont
  {Zhong}}, \bibinfo {author} {\bibfnamefont {G.}~\bibnamefont {Zhang}}, \ and\
  \bibinfo {author} {\bibfnamefont {Y.~W.}\ \bibnamefont {Zhang}},\ }\href@noop
  {} {\bibfield  {journal} {\bibinfo  {journal} {Appl. Phys. Lett.}\ }\textbf
  {\bibinfo {volume} {115}},\ \bibinfo {pages} {063507} (\bibinfo {year}
  {2014})}\BibitemShut {NoStop}%
\bibitem [{\citenamefont {Alofi}\ and\ \citenamefont
  {Srivastava}(2013)}]{alofi13}%
  \BibitemOpen
  \bibfield  {author} {\bibinfo {author} {\bibfnamefont {A.}~\bibnamefont
  {Alofi}}\ and\ \bibinfo {author} {\bibfnamefont {G.~P.}\ \bibnamefont
  {Srivastava}},\ }\href@noop {} {\bibfield  {journal} {\bibinfo  {journal}
  {Phys. Rev. B}\ }\textbf {\bibinfo {volume} {87}},\ \bibinfo {pages} {115421}
  (\bibinfo {year} {2013})}\BibitemShut {NoStop}%
\bibitem [{\citenamefont {Kong}\ \emph {et~al.}(2009)\citenamefont {Kong},
  \citenamefont {Paul}, \citenamefont {Nardelli},\ and\ \citenamefont
  {Kim}}]{kong2009}%
  \BibitemOpen
  \bibfield  {author} {\bibinfo {author} {\bibfnamefont {B.~D.}\ \bibnamefont
  {Kong}}, \bibinfo {author} {\bibfnamefont {S.}~\bibnamefont {Paul}}, \bibinfo
  {author} {\bibfnamefont {M.~B.}\ \bibnamefont {Nardelli}}, \ and\ \bibinfo
  {author} {\bibfnamefont {K.~W.}\ \bibnamefont {Kim}},\ }\href@noop {}
  {\bibfield  {journal} {\bibinfo  {journal} {Phys. Rev. B}\ }\textbf {\bibinfo
  {volume} {80}},\ \bibinfo {pages} {033406} (\bibinfo {year}
  {2009})}\BibitemShut {NoStop}%
\bibitem [{\citenamefont {Aksamija}\ and\ \citenamefont
  {Knezevic}(2011)}]{aksamija11}%
  \BibitemOpen
  \bibfield  {author} {\bibinfo {author} {\bibfnamefont {Z.}~\bibnamefont
  {Aksamija}}\ and\ \bibinfo {author} {\bibfnamefont {I.}~\bibnamefont
  {Knezevic}},\ }\href@noop {} {\bibfield  {journal} {\bibinfo  {journal}
  {Applied Physics Letters}\ }\textbf {\bibinfo {volume} {98}},\ \bibinfo
  {pages} {141919} (\bibinfo {year} {2011})}\BibitemShut {NoStop}%
\bibitem [{\citenamefont {Nika}\ \emph
  {et~al.}(2009{\natexlab{a}})\citenamefont {Nika}, \citenamefont {Pokatilov},
  \citenamefont {Askerov},\ and\ \citenamefont {Balandin}}]{nika2009}%
  \BibitemOpen
  \bibfield  {author} {\bibinfo {author} {\bibfnamefont {D.~L.}\ \bibnamefont
  {Nika}}, \bibinfo {author} {\bibfnamefont {E.~P.}\ \bibnamefont {Pokatilov}},
  \bibinfo {author} {\bibfnamefont {A.~S.}\ \bibnamefont {Askerov}}, \ and\
  \bibinfo {author} {\bibfnamefont {A.~A.}\ \bibnamefont {Balandin}},\
  }\href@noop {} {\bibfield  {journal} {\bibinfo  {journal} {Phys. Rev. B}\
  }\textbf {\bibinfo {volume} {79}},\ \bibinfo {pages} {155413} (\bibinfo
  {year} {2009}{\natexlab{a}})}\BibitemShut {NoStop}%
\bibitem [{\citenamefont {Nika}, \citenamefont {Pokatilov},\ and\ \citenamefont
  {Balandin}(2011)}]{nika11}%
  \BibitemOpen
  \bibfield  {author} {\bibinfo {author} {\bibfnamefont {D.~L.}\ \bibnamefont
  {Nika}}, \bibinfo {author} {\bibfnamefont {E.~P.}\ \bibnamefont {Pokatilov}},
  \ and\ \bibinfo {author} {\bibfnamefont {A.~A.}\ \bibnamefont {Balandin}},\
  }\href@noop {} {\bibfield  {journal} {\bibinfo  {journal} {Physica Status
  Solidi B}\ }\textbf {\bibinfo {volume} {248}},\ \bibinfo {pages} {2609}
  (\bibinfo {year} {2011})}\BibitemShut {NoStop}%
\bibitem [{\citenamefont {Nika}\ and\ \citenamefont {Balandin}(2012)}]{nika12}%
  \BibitemOpen
  \bibfield  {author} {\bibinfo {author} {\bibfnamefont {D.~L.}\ \bibnamefont
  {Nika}}\ and\ \bibinfo {author} {\bibfnamefont {A.~A.}\ \bibnamefont
  {Balandin}},\ }\href@noop {} {\bibfield  {journal} {\bibinfo  {journal} {J.
  Phys.: Condens. Matter}\ }\textbf {\bibinfo {volume} {24}},\ \bibinfo {pages}
  {233203} (\bibinfo {year} {2012})}\BibitemShut {NoStop}%
\bibitem [{\citenamefont {Wei}\ \emph {et~al.}(2014)\citenamefont {Wei},
  \citenamefont {Yang}, \citenamefont {Bi},\ and\ \citenamefont
  {Chen}}]{wei14}%
  \BibitemOpen
  \bibfield  {author} {\bibinfo {author} {\bibfnamefont {Z.}~\bibnamefont
  {Wei}}, \bibinfo {author} {\bibfnamefont {J.}~\bibnamefont {Yang}}, \bibinfo
  {author} {\bibfnamefont {K.}~\bibnamefont {Bi}}, \ and\ \bibinfo {author}
  {\bibfnamefont {Y.}~\bibnamefont {Chen}},\ }\href@noop {} {\bibfield
  {journal} {\bibinfo  {journal} {Journal of Applied Physics}\ }\textbf
  {\bibinfo {volume} {116}},\ \bibinfo {pages} {153503} (\bibinfo {year}
  {2014})}\BibitemShut {NoStop}%
\bibitem [{\citenamefont {Callaway}(1959)}]{callaway59}%
  \BibitemOpen
  \bibfield  {author} {\bibinfo {author} {\bibfnamefont {J.}~\bibnamefont
  {Callaway}},\ }\href@noop {} {\bibfield  {journal} {\bibinfo  {journal}
  {Phys. Rev.}\ }\textbf {\bibinfo {volume} {113}},\ \bibinfo {pages} {1046}
  (\bibinfo {year} {1959})}\BibitemShut {NoStop}%
\bibitem [{\citenamefont {Allen}(2013)}]{allen13}%
  \BibitemOpen
  \bibfield  {author} {\bibinfo {author} {\bibfnamefont {P.~B.}\ \bibnamefont
  {Allen}},\ }\href@noop {} {\bibfield  {journal} {\bibinfo  {journal} {Phys.
  Rev. B}\ }\textbf {\bibinfo {volume} {88}},\ \bibinfo {pages} {144302}
  (\bibinfo {year} {2013})}\BibitemShut {NoStop}%
\bibitem [{\citenamefont {Ma}\ and\ \citenamefont {Luo}(2014)}]{ma14}%
  \BibitemOpen
  \bibfield  {author} {\bibinfo {author} {\bibfnamefont {J.}~\bibnamefont
  {Ma}}\ and\ \bibinfo {author} {\bibfnamefont {W.~L.~X.}\ \bibnamefont
  {Luo}},\ }\href@noop {} {\bibfield  {journal} {\bibinfo  {journal} {Physical
  Review B}\ }\textbf {\bibinfo {volume} {90}},\ \bibinfo {pages} {035203}
  (\bibinfo {year} {2014})}\BibitemShut {NoStop}%
\bibitem [{\citenamefont {Mei}\ \emph {et~al.}(2014)\citenamefont {Mei},
  \citenamefont {Maurer}, \citenamefont {Aksamija},\ and\ \citenamefont
  {Knezevic}}]{mei14}%
  \BibitemOpen
  \bibfield  {author} {\bibinfo {author} {\bibfnamefont {S.}~\bibnamefont
  {Mei}}, \bibinfo {author} {\bibfnamefont {L.~N.}\ \bibnamefont {Maurer}},
  \bibinfo {author} {\bibfnamefont {Z.}~\bibnamefont {Aksamija}}, \ and\
  \bibinfo {author} {\bibfnamefont {I.}~\bibnamefont {Knezevic}},\ }\href@noop
  {} {\bibfield  {journal} {\bibinfo  {journal} {Journ. of Appl. Phys.}\
  }\textbf {\bibinfo {volume} {116}},\ \bibinfo {pages} {164307} (\bibinfo
  {year} {2014})}\BibitemShut {NoStop}%
\bibitem [{\citenamefont {Majee}\ and\ \citenamefont
  {Aksamija}(2016)}]{majee16}%
  \BibitemOpen
  \bibfield  {author} {\bibinfo {author} {\bibfnamefont {A.~K.}\ \bibnamefont
  {Majee}}\ and\ \bibinfo {author} {\bibfnamefont {Z.}~\bibnamefont
  {Aksamija}},\ }\href@noop {} {\bibfield  {journal} {\bibinfo  {journal}
  {Phys. Rev. B}\ }\textbf {\bibinfo {volume} {93}},\ \bibinfo {pages} {235423}
  (\bibinfo {year} {2016})}\BibitemShut {NoStop}%
\bibitem [{\citenamefont {Guo}\ and\ \citenamefont {Wang}(2017)}]{guo17}%
  \BibitemOpen
  \bibfield  {author} {\bibinfo {author} {\bibfnamefont {Y.}~\bibnamefont
  {Guo}}\ and\ \bibinfo {author} {\bibfnamefont {M.}~\bibnamefont {Wang}},\
  }\href@noop {} {\bibfield  {journal} {\bibinfo  {journal} {Phys. Rev. B}\
  }\textbf {\bibinfo {volume} {96}},\ \bibinfo {pages} {134312} (\bibinfo
  {year} {2017})}\BibitemShut {NoStop}%
\bibitem [{\citenamefont {Fugallo}\ \emph {et~al.}(2014)\citenamefont
  {Fugallo}, \citenamefont {Cepellotti}, \citenamefont {Paulatto},
  \citenamefont {Lazzeri}, \citenamefont {Marzari},\ and\ \citenamefont
  {Mauri}}]{fugallo14}%
  \BibitemOpen
  \bibfield  {author} {\bibinfo {author} {\bibfnamefont {G.}~\bibnamefont
  {Fugallo}}, \bibinfo {author} {\bibfnamefont {A.}~\bibnamefont {Cepellotti}},
  \bibinfo {author} {\bibfnamefont {L.}~\bibnamefont {Paulatto}}, \bibinfo
  {author} {\bibfnamefont {M.}~\bibnamefont {Lazzeri}}, \bibinfo {author}
  {\bibfnamefont {N.}~\bibnamefont {Marzari}}, \ and\ \bibinfo {author}
  {\bibfnamefont {F.}~\bibnamefont {Mauri}},\ }\href@noop {} {\bibfield
  {journal} {\bibinfo  {journal} {Nano Lett.}\ }\textbf {\bibinfo {volume}
  {14}},\ \bibinfo {pages} {6109} (\bibinfo {year} {2014})}\BibitemShut
  {NoStop}%
\bibitem [{\citenamefont {Polanco}\ and\ \citenamefont
  {Lindsay}(2018)}]{polanco18}%
  \BibitemOpen
  \bibfield  {author} {\bibinfo {author} {\bibfnamefont {C.~A.}\ \bibnamefont
  {Polanco}}\ and\ \bibinfo {author} {\bibfnamefont {L.}~\bibnamefont
  {Lindsay}},\ }\href@noop {} {\bibfield  {journal} {\bibinfo  {journal} {Phys.
  Rev. B}\ }\textbf {\bibinfo {volume} {97}},\ \bibinfo {pages} {014303}
  (\bibinfo {year} {2018})}\BibitemShut {NoStop}%
\bibitem [{\citenamefont {Li}\ \emph {et~al.}(2014{\natexlab{a}})\citenamefont
  {Li}, \citenamefont {Carrete}, \citenamefont {Katcho},\ and\ \citenamefont
  {Mingo}}]{ShengBTE}%
  \BibitemOpen
  \bibfield  {author} {\bibinfo {author} {\bibfnamefont {W.}~\bibnamefont
  {Li}}, \bibinfo {author} {\bibfnamefont {J.}~\bibnamefont {Carrete}},
  \bibinfo {author} {\bibfnamefont {N.~A.}\ \bibnamefont {Katcho}}, \ and\
  \bibinfo {author} {\bibfnamefont {N.}~\bibnamefont {Mingo}},\ }\href@noop {}
  {\bibfield  {journal} {\bibinfo  {journal} {Comp. Phys. Commun.}\ }\textbf
  {\bibinfo {volume} {185}},\ \bibinfo {pages} {1747} (\bibinfo {year}
  {2014}{\natexlab{a}})}\BibitemShut {NoStop}%
\bibitem [{\citenamefont {Anno}\ \emph {et~al.}(2017)\citenamefont {Anno},
  \citenamefont {Imakita}, \citenamefont {Takei}, \citenamefont {Akita},\ and\
  \citenamefont {Arie}}]{anno17}%
  \BibitemOpen
  \bibfield  {author} {\bibinfo {author} {\bibfnamefont {Y.}~\bibnamefont
  {Anno}}, \bibinfo {author} {\bibfnamefont {Y.}~\bibnamefont {Imakita}},
  \bibinfo {author} {\bibfnamefont {K.}~\bibnamefont {Takei}}, \bibinfo
  {author} {\bibfnamefont {S.}~\bibnamefont {Akita}}, \ and\ \bibinfo {author}
  {\bibfnamefont {T.}~\bibnamefont {Arie}},\ }\href@noop {} {\bibfield
  {journal} {\bibinfo  {journal} {2D Materials}\ }\textbf {\bibinfo {volume}
  {4}},\ \bibinfo {pages} {025019} (\bibinfo {year} {2017})}\BibitemShut
  {NoStop}%
\bibitem [{\citenamefont {Jorio}\ \emph {et~al.}(2011)\citenamefont {Jorio},
  \citenamefont {Dresselhaus}, \citenamefont {Saito},\ and\ \citenamefont
  {Dresselhaus}}]{jorio11}%
  \BibitemOpen
  \bibfield  {author} {\bibinfo {author} {\bibfnamefont {A.}~\bibnamefont
  {Jorio}}, \bibinfo {author} {\bibfnamefont {M.~S.}\ \bibnamefont
  {Dresselhaus}}, \bibinfo {author} {\bibfnamefont {R.}~\bibnamefont {Saito}},
  \ and\ \bibinfo {author} {\bibfnamefont {G.}~\bibnamefont {Dresselhaus}},\
  }\href@noop {} {\emph {\bibinfo {title} {Raman Spectroscopy in Graphene
  Related Systems}}}\ (\bibinfo  {publisher} {Wiley},\ \bibinfo {address}
  {Weinheim},\ \bibinfo {year} {2011})\BibitemShut {NoStop}%
\bibitem [{\citenamefont {Giannozzi}\ and\ \citenamefont
  {et~al.}(2009)}]{giannozzi09}%
  \BibitemOpen
  \bibfield  {author} {\bibinfo {author} {\bibfnamefont {P.}~\bibnamefont
  {Giannozzi}}\ and\ \bibinfo {author} {\bibnamefont {et~al.}},\ }\href@noop {}
  {\bibfield  {journal} {\bibinfo  {journal} {J. Phys. Condens. Matter}\
  }\textbf {\bibinfo {volume} {21}},\ \bibinfo {pages} {395502} (\bibinfo
  {year} {2009})}\BibitemShut {NoStop}%
\bibitem [{\citenamefont {Rappe}\ \emph {et~al.}(1990)\citenamefont {Rappe},
  \citenamefont {Rabe}, \citenamefont {Kaxiras},\ and\ \citenamefont
  {Joannopoulos}}]{rrkj90}%
  \BibitemOpen
  \bibfield  {author} {\bibinfo {author} {\bibfnamefont {A.~M.}\ \bibnamefont
  {Rappe}}, \bibinfo {author} {\bibfnamefont {K.~M.}\ \bibnamefont {Rabe}},
  \bibinfo {author} {\bibfnamefont {E.}~\bibnamefont {Kaxiras}}, \ and\
  \bibinfo {author} {\bibfnamefont {J.~D.}\ \bibnamefont {Joannopoulos}},\
  }\href@noop {} {\bibfield  {journal} {\bibinfo  {journal} {Phys. Rev. B}\
  }\textbf {\bibinfo {volume} {41}},\ \bibinfo {pages} {1227} (\bibinfo {year}
  {1990})}\BibitemShut {NoStop}%
\bibitem [{\citenamefont {Grimme}(2006)}]{grimme1}%
  \BibitemOpen
  \bibfield  {author} {\bibinfo {author} {\bibfnamefont {S.}~\bibnamefont
  {Grimme}},\ }\href@noop {} {\bibfield  {journal} {\bibinfo  {journal} {J.
  Comp. Chem.}\ }\textbf {\bibinfo {volume} {27}},\ \bibinfo {pages} {1787}
  (\bibinfo {year} {2006})}\BibitemShut {NoStop}%
\bibitem [{\citenamefont {Monkhorst}\ and\ \citenamefont {Pack}(1976)}]{mp76}%
  \BibitemOpen
  \bibfield  {author} {\bibinfo {author} {\bibfnamefont {H.~J.}\ \bibnamefont
  {Monkhorst}}\ and\ \bibinfo {author} {\bibfnamefont {J.~D.}\ \bibnamefont
  {Pack}},\ }\href@noop {} {\bibfield  {journal} {\bibinfo  {journal} {Phys.
  Rev. B}\ }\textbf {\bibinfo {volume} {13}},\ \bibinfo {pages} {5188}
  (\bibinfo {year} {1976})}\BibitemShut {NoStop}%
\bibitem [{\citenamefont {Baroni}, \citenamefont {Giannozzi},\ and\
  \citenamefont {Testa}(1987)}]{dfpt87}%
  \BibitemOpen
  \bibfield  {author} {\bibinfo {author} {\bibfnamefont {S.}~\bibnamefont
  {Baroni}}, \bibinfo {author} {\bibfnamefont {P.}~\bibnamefont {Giannozzi}}, \
  and\ \bibinfo {author} {\bibfnamefont {A.}~\bibnamefont {Testa}},\
  }\href@noop {} {\bibfield  {journal} {\bibinfo  {journal} {Physical Review
  Letters}\ }\textbf {\bibinfo {volume} {58}},\ \bibinfo {pages} {1861}
  (\bibinfo {year} {1987})}\BibitemShut {NoStop}%
\bibitem [{\citenamefont {Klemens}(1958)}]{klemens58}%
  \BibitemOpen
  \bibfield  {author} {\bibinfo {author} {\bibfnamefont {P.~G.}\ \bibnamefont
  {Klemens}},\ }in\ \href@noop {} {\emph {\bibinfo {booktitle} {Solid State
  Physics: Advances in Research and Applications}}},\ Vol.~\bibinfo {volume}
  {7},\ \bibinfo {editor} {edited by\ \bibinfo {editor} {\bibfnamefont
  {F.}~\bibnamefont {Seitz}}\ and\ \bibinfo {editor} {\bibfnamefont
  {D.}~\bibnamefont {Turnbull}}}\ (\bibinfo  {publisher} {Academic},\ \bibinfo
  {address} {New York},\ \bibinfo {year} {1958})\ p.~\bibinfo {pages}
  {1}\BibitemShut {NoStop}%
\bibitem [{\citenamefont {D'Souza}\ and\ \citenamefont
  {Mukherjee}(2017{\natexlab{b}})}]{RSBN17}%
  \BibitemOpen
  \bibfield  {author} {\bibinfo {author} {\bibfnamefont {R.}~\bibnamefont
  {D'Souza}}\ and\ \bibinfo {author} {\bibfnamefont {S.}~\bibnamefont
  {Mukherjee}},\ }\href@noop {} {\bibfield  {journal} {\bibinfo  {journal}
  {Phys. Rev. B}\ }\textbf {\bibinfo {volume} {96}},\ \bibinfo {pages} {205422}
  (\bibinfo {year} {2017}{\natexlab{b}})}\BibitemShut {NoStop}%
\bibitem [{\citenamefont {Ashcroft}\ and\ \citenamefont
  {Mermin}(1976)}]{ashcroft}%
  \BibitemOpen
  \bibfield  {author} {\bibinfo {author} {\bibfnamefont {N.~W.}\ \bibnamefont
  {Ashcroft}}\ and\ \bibinfo {author} {\bibfnamefont {N.~D.}\ \bibnamefont
  {Mermin}},\ }\href@noop {} {\emph {\bibinfo {title} {Solid State Physics}}}\
  (\bibinfo  {publisher} {Holt, Reinehart and Winston, New York},\ \bibinfo
  {year} {1976})\BibitemShut {NoStop}%
\bibitem [{\citenamefont {Klemens}\ and\ \citenamefont
  {Pedraza}(1994)}]{klemens94}%
  \BibitemOpen
  \bibfield  {author} {\bibinfo {author} {\bibfnamefont {P.~G.}\ \bibnamefont
  {Klemens}}\ and\ \bibinfo {author} {\bibfnamefont {D.~F.}\ \bibnamefont
  {Pedraza}},\ }\href@noop {} {\bibfield  {journal} {\bibinfo  {journal}
  {Carbon}\ }\textbf {\bibinfo {volume} {32}},\ \bibinfo {pages} {735}
  (\bibinfo {year} {1994})}\BibitemShut {NoStop}%
\bibitem [{\citenamefont {Malard}\ \emph {et~al.}(2009)\citenamefont {Malard},
  \citenamefont {Pimenta}, \citenamefont {Dresselhaus},\ and\ \citenamefont
  {Dresselhaus}}]{dresselhaus09}%
  \BibitemOpen
  \bibfield  {author} {\bibinfo {author} {\bibfnamefont {L.}~\bibnamefont
  {Malard}}, \bibinfo {author} {\bibfnamefont {M.}~\bibnamefont {Pimenta}},
  \bibinfo {author} {\bibfnamefont {G.}~\bibnamefont {Dresselhaus}}, \ and\
  \bibinfo {author} {\bibfnamefont {M.}~\bibnamefont {Dresselhaus}},\
  }\href@noop {} {\bibfield  {journal} {\bibinfo  {journal} {Physics Reports}\
  }\textbf {\bibinfo {volume} {473}},\ \bibinfo {pages} {51} (\bibinfo {year}
  {2009})}\BibitemShut {NoStop}%
\bibitem [{\citenamefont {Yanagisawa}\ \emph {et~al.}(2005)\citenamefont
  {Yanagisawa}, \citenamefont {Tanaka}, \citenamefont {Ishida}, \citenamefont
  {Matsue}, \citenamefont {Rokuta}, \citenamefont {Otani},\ and\ \citenamefont
  {Oshima}}]{yanagisawa05}%
  \BibitemOpen
  \bibfield  {author} {\bibinfo {author} {\bibfnamefont {H.}~\bibnamefont
  {Yanagisawa}}, \bibinfo {author} {\bibfnamefont {T.}~\bibnamefont {Tanaka}},
  \bibinfo {author} {\bibfnamefont {Y.}~\bibnamefont {Ishida}}, \bibinfo
  {author} {\bibfnamefont {M.}~\bibnamefont {Matsue}}, \bibinfo {author}
  {\bibfnamefont {E.}~\bibnamefont {Rokuta}}, \bibinfo {author} {\bibfnamefont
  {S.}~\bibnamefont {Otani}}, \ and\ \bibinfo {author} {\bibfnamefont
  {C.}~\bibnamefont {Oshima}},\ }\href@noop {} {\bibfield  {journal} {\bibinfo
  {journal} {Surf. Interface Anal.}\ }\textbf {\bibinfo {volume} {37}},\
  \bibinfo {pages} {133} (\bibinfo {year} {2005})}\BibitemShut {NoStop}%
\bibitem [{\citenamefont {Lui}\ and\ \citenamefont {Heinz}(2013)}]{lui13}%
  \BibitemOpen
  \bibfield  {author} {\bibinfo {author} {\bibfnamefont {C.~H.}\ \bibnamefont
  {Lui}}\ and\ \bibinfo {author} {\bibfnamefont {T.~F.}\ \bibnamefont
  {Heinz}},\ }\href@noop {} {\bibfield  {journal} {\bibinfo  {journal} {Phys.
  Rev. B}\ }\textbf {\bibinfo {volume} {87}},\ \bibinfo {pages} {121404}
  (\bibinfo {year} {2013})}\BibitemShut {NoStop}%
\bibitem [{\citenamefont {Tan}\ \emph {et~al.}(2012)\citenamefont {Tan},
  \citenamefont {Han}, \citenamefont {Zhao}, \citenamefont {Wu}, \citenamefont
  {Chang}, \citenamefont {Wang}, \citenamefont {Wang}, \citenamefont {Bonini},
  \citenamefont {Marzari}, \citenamefont {Pugno}, \citenamefont {Savini},
  \citenamefont {Lombardo},\ and\ \citenamefont {Ferrari}}]{tan12}%
  \BibitemOpen
  \bibfield  {author} {\bibinfo {author} {\bibfnamefont {P.~H.}\ \bibnamefont
  {Tan}}, \bibinfo {author} {\bibfnamefont {W.~P.}\ \bibnamefont {Han}},
  \bibinfo {author} {\bibfnamefont {W.~J.}\ \bibnamefont {Zhao}}, \bibinfo
  {author} {\bibfnamefont {Z.~H.}\ \bibnamefont {Wu}}, \bibinfo {author}
  {\bibfnamefont {K.}~\bibnamefont {Chang}}, \bibinfo {author} {\bibfnamefont
  {H.}~\bibnamefont {Wang}}, \bibinfo {author} {\bibfnamefont {Y.~F.}\
  \bibnamefont {Wang}}, \bibinfo {author} {\bibfnamefont {N.}~\bibnamefont
  {Bonini}}, \bibinfo {author} {\bibfnamefont {N.}~\bibnamefont {Marzari}},
  \bibinfo {author} {\bibfnamefont {N.}~\bibnamefont {Pugno}}, \bibinfo
  {author} {\bibfnamefont {G.}~\bibnamefont {Savini}}, \bibinfo {author}
  {\bibfnamefont {A.}~\bibnamefont {Lombardo}}, \ and\ \bibinfo {author}
  {\bibfnamefont {A.~C.}\ \bibnamefont {Ferrari}},\ }\href@noop {} {\bibfield
  {journal} {\bibinfo  {journal} {Nature materials}\ }\textbf {\bibinfo
  {volume} {11}},\ \bibinfo {pages} {294} (\bibinfo {year} {2012})}\BibitemShut
  {NoStop}%
\bibitem [{\citenamefont {Mohiuddin}\ \emph {et~al.}(2009)\citenamefont
  {Mohiuddin}, \citenamefont {Lombardo}, \citenamefont {Nair}, \citenamefont
  {Bonetti}, \citenamefont {Savini}, \citenamefont {Jalil}, \citenamefont
  {Bonini}, \citenamefont {Basko}, \citenamefont {Galiotis}, \citenamefont
  {Marzari}, \citenamefont {Novoselov}, \citenamefont {Geim},\ and\
  \citenamefont {Ferrari}}]{mohiuddin09}%
  \BibitemOpen
  \bibfield  {author} {\bibinfo {author} {\bibfnamefont {T.~M.~G.}\
  \bibnamefont {Mohiuddin}}, \bibinfo {author} {\bibfnamefont {A.}~\bibnamefont
  {Lombardo}}, \bibinfo {author} {\bibfnamefont {R.~R.}\ \bibnamefont {Nair}},
  \bibinfo {author} {\bibfnamefont {A.}~\bibnamefont {Bonetti}}, \bibinfo
  {author} {\bibfnamefont {G.}~\bibnamefont {Savini}}, \bibinfo {author}
  {\bibfnamefont {R.}~\bibnamefont {Jalil}}, \bibinfo {author} {\bibfnamefont
  {N.}~\bibnamefont {Bonini}}, \bibinfo {author} {\bibfnamefont {D.~M.}\
  \bibnamefont {Basko}}, \bibinfo {author} {\bibfnamefont {C.}~\bibnamefont
  {Galiotis}}, \bibinfo {author} {\bibfnamefont {N.}~\bibnamefont {Marzari}},
  \bibinfo {author} {\bibfnamefont {K.~S.}\ \bibnamefont {Novoselov}}, \bibinfo
  {author} {\bibfnamefont {A.~K.}\ \bibnamefont {Geim}}, \ and\ \bibinfo
  {author} {\bibfnamefont {A.~C.}\ \bibnamefont {Ferrari}},\ }\href@noop {}
  {\bibfield  {journal} {\bibinfo  {journal} {Physical Review B}\ }\textbf
  {\bibinfo {volume} {79}},\ \bibinfo {pages} {205433} (\bibinfo {year}
  {2009})}\BibitemShut {NoStop}%
\bibitem [{\citenamefont {Mounet}\ and\ \citenamefont
  {Marzari}(2005)}]{marzari05}%
  \BibitemOpen
  \bibfield  {author} {\bibinfo {author} {\bibfnamefont {N.}~\bibnamefont
  {Mounet}}\ and\ \bibinfo {author} {\bibfnamefont {N.}~\bibnamefont
  {Marzari}},\ }\href@noop {} {\bibfield  {journal} {\bibinfo  {journal}
  {Physical Review B}\ }\textbf {\bibinfo {volume} {71}},\ \bibinfo {pages}
  {205214} (\bibinfo {year} {2005})}\BibitemShut {NoStop}%
\bibitem [{\citenamefont {Pettes}\ \emph {et~al.}(2011)\citenamefont {Pettes},
  \citenamefont {Jo}, \citenamefont {Yao},\ and\ \citenamefont
  {Shi}}]{pettes11}%
  \BibitemOpen
  \bibfield  {author} {\bibinfo {author} {\bibfnamefont {M.~T.}\ \bibnamefont
  {Pettes}}, \bibinfo {author} {\bibfnamefont {I.}~\bibnamefont {Jo}}, \bibinfo
  {author} {\bibfnamefont {Z.}~\bibnamefont {Yao}}, \ and\ \bibinfo {author}
  {\bibfnamefont {L.}~\bibnamefont {Shi}},\ }\href@noop {} {\bibfield
  {journal} {\bibinfo  {journal} {Nano Lett.}\ }\textbf {\bibinfo {volume}
  {11}},\ \bibinfo {pages} {1195} (\bibinfo {year} {2011})}\BibitemShut
  {NoStop}%
\bibitem [{\citenamefont {Xu}\ \emph {et~al.}(2014)\citenamefont {Xu},
  \citenamefont {Pereira}, \citenamefont {Wang}, \citenamefont {Wu},
  \citenamefont {Zhang}, \citenamefont {Zhao}, \citenamefont {Bae},
  \citenamefont {Bui}, \citenamefont {Xie}, \citenamefont {Thong},
  \citenamefont {Hong}, \citenamefont {Loh}, \citenamefont {Donadio},
  \citenamefont {Li},\ and\ \citenamefont {\"{O}zyilmaz}}]{xu14}%
  \BibitemOpen
  \bibfield  {author} {\bibinfo {author} {\bibfnamefont {X.}~\bibnamefont
  {Xu}}, \bibinfo {author} {\bibfnamefont {L.~F.}\ \bibnamefont {Pereira}},
  \bibinfo {author} {\bibfnamefont {Y.}~\bibnamefont {Wang}}, \bibinfo {author}
  {\bibfnamefont {J.}~\bibnamefont {Wu}}, \bibinfo {author} {\bibfnamefont
  {K.}~\bibnamefont {Zhang}}, \bibinfo {author} {\bibfnamefont
  {X.}~\bibnamefont {Zhao}}, \bibinfo {author} {\bibfnamefont {S.}~\bibnamefont
  {Bae}}, \bibinfo {author} {\bibfnamefont {C.~T.}\ \bibnamefont {Bui}},
  \bibinfo {author} {\bibfnamefont {R.}~\bibnamefont {Xie}}, \bibinfo {author}
  {\bibfnamefont {J.~T.}\ \bibnamefont {Thong}}, \bibinfo {author}
  {\bibfnamefont {B.~H.}\ \bibnamefont {Hong}}, \bibinfo {author}
  {\bibfnamefont {K.~P.}\ \bibnamefont {Loh}}, \bibinfo {author} {\bibfnamefont
  {D.}~\bibnamefont {Donadio}}, \bibinfo {author} {\bibfnamefont
  {B.}~\bibnamefont {Li}}, \ and\ \bibinfo {author} {\bibfnamefont
  {B.}~\bibnamefont {\"{O}zyilmaz}},\ }\href@noop {} {\bibfield  {journal}
  {\bibinfo  {journal} {Nature Communications}\ }\textbf {\bibinfo {volume}
  {5}},\ \bibinfo {pages} {3689} (\bibinfo {year} {2014})}\BibitemShut
  {NoStop}%
\bibitem [{\citenamefont {Li}\ \emph {et~al.}(2014{\natexlab{b}})\citenamefont
  {Li}, \citenamefont {Ying}, \citenamefont {Chen}, \citenamefont {Nika},
  \citenamefont {Cocemasov}, \citenamefont {Cai}, \citenamefont {Balandin},\
  and\ \citenamefont {Chen}}]{hongyang2014}%
  \BibitemOpen
  \bibfield  {author} {\bibinfo {author} {\bibfnamefont {H.}~\bibnamefont
  {Li}}, \bibinfo {author} {\bibfnamefont {H.}~\bibnamefont {Ying}}, \bibinfo
  {author} {\bibfnamefont {X.}~\bibnamefont {Chen}}, \bibinfo {author}
  {\bibfnamefont {D.~L.}\ \bibnamefont {Nika}}, \bibinfo {author}
  {\bibfnamefont {A.~I.}\ \bibnamefont {Cocemasov}}, \bibinfo {author}
  {\bibfnamefont {W.}~\bibnamefont {Cai}}, \bibinfo {author} {\bibfnamefont
  {A.~A.}\ \bibnamefont {Balandin}}, \ and\ \bibinfo {author} {\bibfnamefont
  {S.}~\bibnamefont {Chen}},\ }\href@noop {} {\bibfield  {journal} {\bibinfo
  {journal} {Nanoscale}\ }\textbf {\bibinfo {volume} {6}},\ \bibinfo {pages}
  {13402} (\bibinfo {year} {2014}{\natexlab{b}})}\BibitemShut {NoStop}%
\bibitem [{\citenamefont {Lindsay}\ and\ \citenamefont
  {Broido}(2012)}]{lindsay12}%
  \BibitemOpen
  \bibfield  {author} {\bibinfo {author} {\bibfnamefont {L.}~\bibnamefont
  {Lindsay}}\ and\ \bibinfo {author} {\bibfnamefont {D.~A.}\ \bibnamefont
  {Broido}},\ }\href@noop {} {\bibfield  {journal} {\bibinfo  {journal} {Phys.
  Rev. B}\ }\textbf {\bibinfo {volume} {85}},\ \bibinfo {pages} {035436}
  (\bibinfo {year} {2012})}\BibitemShut {NoStop}%
\bibitem [{\citenamefont {Berger}\ \emph {et~al.}(2006)\citenamefont {Berger},
  \citenamefont {Song}, \citenamefont {Li}, \citenamefont {Wu}, \citenamefont
  {Brown}, \citenamefont {Naud}, \citenamefont {Mayou}, \citenamefont {Li},
  \citenamefont {Hass}, \citenamefont {Marchenkov}, \citenamefont {Conrad},
  \citenamefont {First},\ and\ \citenamefont {de~Heer}}]{berger06}%
  \BibitemOpen
  \bibfield  {author} {\bibinfo {author} {\bibfnamefont {C.}~\bibnamefont
  {Berger}}, \bibinfo {author} {\bibfnamefont {Z.}~\bibnamefont {Song}},
  \bibinfo {author} {\bibfnamefont {X.}~\bibnamefont {Li}}, \bibinfo {author}
  {\bibfnamefont {X.}~\bibnamefont {Wu}}, \bibinfo {author} {\bibfnamefont
  {N.}~\bibnamefont {Brown}}, \bibinfo {author} {\bibfnamefont
  {C.}~\bibnamefont {Naud}}, \bibinfo {author} {\bibfnamefont {D.}~\bibnamefont
  {Mayou}}, \bibinfo {author} {\bibfnamefont {T.}~\bibnamefont {Li}}, \bibinfo
  {author} {\bibfnamefont {J.}~\bibnamefont {Hass}}, \bibinfo {author}
  {\bibfnamefont {A.~N.}\ \bibnamefont {Marchenkov}}, \bibinfo {author}
  {\bibfnamefont {E.~H.}\ \bibnamefont {Conrad}}, \bibinfo {author}
  {\bibfnamefont {P.~N.}\ \bibnamefont {First}}, \ and\ \bibinfo {author}
  {\bibfnamefont {W.~A.}\ \bibnamefont {de~Heer}},\ }\href@noop {} {\bibfield
  {journal} {\bibinfo  {journal} {Science}\ }\textbf {\bibinfo {volume}
  {312}},\ \bibinfo {pages} {1191} (\bibinfo {year} {2006})}\BibitemShut
  {NoStop}%
\bibitem [{\citenamefont {Fang}\ \emph {et~al.}(2015)\citenamefont {Fang},
  \citenamefont {Yu}, \citenamefont {Zheng}, \citenamefont {Jin}, \citenamefont
  {Wang},\ and\ \citenamefont {Cao}}]{fang15}%
  \BibitemOpen
  \bibfield  {author} {\bibinfo {author} {\bibfnamefont {X.-Y.}\ \bibnamefont
  {Fang}}, \bibinfo {author} {\bibfnamefont {X.-X.}\ \bibnamefont {Yu}},
  \bibinfo {author} {\bibfnamefont {H.-M.}\ \bibnamefont {Zheng}}, \bibinfo
  {author} {\bibfnamefont {H.-B.}\ \bibnamefont {Jin}}, \bibinfo {author}
  {\bibfnamefont {L.}~\bibnamefont {Wang}}, \ and\ \bibinfo {author}
  {\bibfnamefont {M.-S.}\ \bibnamefont {Cao}},\ }\href@noop {} {\bibfield
  {journal} {\bibinfo  {journal} {Phys. Lett. A}\ }\textbf {\bibinfo {volume}
  {379}},\ \bibinfo {pages} {2245} (\bibinfo {year} {2015})}\BibitemShut
  {NoStop}%
\bibitem [{\citenamefont {Durczewski}\ and\ \citenamefont
  {Ausloos}(2000)}]{durczewski2000}%
  \BibitemOpen
  \bibfield  {author} {\bibinfo {author} {\bibfnamefont {K.}~\bibnamefont
  {Durczewski}}\ and\ \bibinfo {author} {\bibfnamefont {M.}~\bibnamefont
  {Ausloos}},\ }\href@noop {} {\bibfield  {journal} {\bibinfo  {journal} {Phys.
  Rev. B}\ }\textbf {\bibinfo {volume} {97}},\ \bibinfo {pages} {5303}
  (\bibinfo {year} {2000})}\BibitemShut {NoStop}%
\bibitem [{\citenamefont {Zahedifar}\ and\ \citenamefont
  {Kratzer}(2018)}]{zahedifar18}%
  \BibitemOpen
  \bibfield  {author} {\bibinfo {author} {\bibfnamefont {M.}~\bibnamefont
  {Zahedifar}}\ and\ \bibinfo {author} {\bibfnamefont {P.}~\bibnamefont
  {Kratzer}},\ }\href@noop {} {\bibfield  {journal} {\bibinfo  {journal} {Phys.
  Rev. B}\ }\textbf {\bibinfo {volume} {97}},\ \bibinfo {pages} {035204}
  (\bibinfo {year} {2018})}\BibitemShut {NoStop}%
\bibitem [{\citenamefont {Madsen}\ and\ \citenamefont
  {Singh}(2006)}]{boltztrap1}%
  \BibitemOpen
  \bibfield  {author} {\bibinfo {author} {\bibfnamefont {G.~K.~H.}\
  \bibnamefont {Madsen}}\ and\ \bibinfo {author} {\bibfnamefont {D.~J.}\
  \bibnamefont {Singh}},\ }\href@noop {} {\bibfield  {journal} {\bibinfo
  {journal} {Computer Physics Communication}\ }\textbf {\bibinfo {volume}
  {175}},\ \bibinfo {pages} {67} (\bibinfo {year} {2006})}\BibitemShut
  {NoStop}%
\bibitem [{\citenamefont {Schulz}, \citenamefont {Allen},\ and\ \citenamefont
  {Trivedi}(1992)}]{schulz92}%
  \BibitemOpen
  \bibfield  {author} {\bibinfo {author} {\bibfnamefont {W.}~\bibnamefont
  {Schulz}}, \bibinfo {author} {\bibfnamefont {P.}~\bibnamefont {Allen}}, \
  and\ \bibinfo {author} {\bibfnamefont {N.}~\bibnamefont {Trivedi}},\
  }\href@noop {} {\bibfield  {journal} {\bibinfo  {journal} {Phys. Rev. B}\
  }\textbf {\bibinfo {volume} {45}},\ \bibinfo {pages} {10886} (\bibinfo {year}
  {1992})}\BibitemShut {NoStop}%
\bibitem [{\citenamefont {Zuev}, \citenamefont {Chang},\ and\ \citenamefont
  {Kim}(2009{\natexlab{b}})}]{kim}%
  \BibitemOpen
  \bibfield  {author} {\bibinfo {author} {\bibfnamefont {Y.}~\bibnamefont
  {Zuev}}, \bibinfo {author} {\bibfnamefont {W.}~\bibnamefont {Chang}}, \ and\
  \bibinfo {author} {\bibfnamefont {P.}~\bibnamefont {Kim}},\ }\href@noop {}
  {\bibfield  {journal} {\bibinfo  {journal} {Physical Review Letters}\
  }\textbf {\bibinfo {volume} {102}},\ \bibinfo {pages} {096807} (\bibinfo
  {year} {2009}{\natexlab{b}})}\BibitemShut {NoStop}%
\bibitem [{\citenamefont {Bolotin}\ \emph
  {et~al.}(2008{\natexlab{b}})\citenamefont {Bolotin}, \citenamefont {Sikes},
  \citenamefont {Hone}, \citenamefont {Stormer},\ and\ \citenamefont
  {Kim}}]{kim2}%
  \BibitemOpen
  \bibfield  {author} {\bibinfo {author} {\bibfnamefont {K.~I.}\ \bibnamefont
  {Bolotin}}, \bibinfo {author} {\bibfnamefont {K.~J.}\ \bibnamefont {Sikes}},
  \bibinfo {author} {\bibfnamefont {J.}~\bibnamefont {Hone}}, \bibinfo {author}
  {\bibfnamefont {H.~L.}\ \bibnamefont {Stormer}}, \ and\ \bibinfo {author}
  {\bibfnamefont {P.}~\bibnamefont {Kim}},\ }\href@noop {} {\bibfield
  {journal} {\bibinfo  {journal} {Physical Review Letters}\ }\textbf {\bibinfo
  {volume} {101}},\ \bibinfo {pages} {096802} (\bibinfo {year}
  {2008}{\natexlab{b}})}\BibitemShut {NoStop}%
\bibitem [{\citenamefont {Morozov}\ \emph {et~al.}(2008)\citenamefont
  {Morozov}, \citenamefont {Novoselov}, \citenamefont {Katsnelson},
  \citenamefont {Schedin}, \citenamefont {Elias}, \citenamefont {Jaszczak},\
  and\ \citenamefont {Geim}}]{morozov08}%
  \BibitemOpen
  \bibfield  {author} {\bibinfo {author} {\bibfnamefont {S.~V.}\ \bibnamefont
  {Morozov}}, \bibinfo {author} {\bibfnamefont {K.~S.}\ \bibnamefont
  {Novoselov}}, \bibinfo {author} {\bibfnamefont {M.~I.}\ \bibnamefont
  {Katsnelson}}, \bibinfo {author} {\bibfnamefont {F.}~\bibnamefont {Schedin}},
  \bibinfo {author} {\bibfnamefont {D.~C.}\ \bibnamefont {Elias}}, \bibinfo
  {author} {\bibfnamefont {J.~A.}\ \bibnamefont {Jaszczak}}, \ and\ \bibinfo
  {author} {\bibfnamefont {A.~K.}\ \bibnamefont {Geim}},\ }\href@noop {}
  {\bibfield  {journal} {\bibinfo  {journal} {Phys. Rev. Lett.}\ }\textbf
  {\bibinfo {volume} {100}},\ \bibinfo {pages} {016602} (\bibinfo {year}
  {2008})}\BibitemShut {NoStop}%
\bibitem [{\citenamefont {Araujo}, \citenamefont {Terrones},\ and\
  \citenamefont {Dresselhaus}(2012)}]{araujo12}%
  \BibitemOpen
  \bibfield  {author} {\bibinfo {author} {\bibfnamefont {P.~T.}\ \bibnamefont
  {Araujo}}, \bibinfo {author} {\bibfnamefont {M.}~\bibnamefont {Terrones}}, \
  and\ \bibinfo {author} {\bibfnamefont {M.~S.}\ \bibnamefont {Dresselhaus}},\
  }\href@noop {} {\bibfield  {journal} {\bibinfo  {journal} {Mater. Today}\
  }\textbf {\bibinfo {volume} {15}},\ \bibinfo {pages} {98} (\bibinfo {year}
  {2012})}\BibitemShut {NoStop}%
\bibitem [{\citenamefont {Genc}\ \emph {et~al.}(2017)\citenamefont {Genc},
  \citenamefont {Alas}, \citenamefont {Harputlu}, \citenamefont {Repp},
  \citenamefont {Kremer}, \citenamefont {Castellano}, \citenamefont {Colak},
  \citenamefont {Ocakoglu},\ and\ \citenamefont {Erdem}}]{genc17}%
  \BibitemOpen
  \bibfield  {author} {\bibinfo {author} {\bibfnamefont {R.}~\bibnamefont
  {Genc}}, \bibinfo {author} {\bibfnamefont {M.~O.}\ \bibnamefont {Alas}},
  \bibinfo {author} {\bibfnamefont {E.}~\bibnamefont {Harputlu}}, \bibinfo
  {author} {\bibfnamefont {S.}~\bibnamefont {Repp}}, \bibinfo {author}
  {\bibfnamefont {N.}~\bibnamefont {Kremer}}, \bibinfo {author} {\bibfnamefont
  {M.}~\bibnamefont {Castellano}}, \bibinfo {author} {\bibfnamefont {S.~G.}\
  \bibnamefont {Colak}}, \bibinfo {author} {\bibfnamefont {K.}~\bibnamefont
  {Ocakoglu}}, \ and\ \bibinfo {author} {\bibfnamefont {E.}~\bibnamefont
  {Erdem}},\ }\href@noop {} {\bibfield  {journal} {\bibinfo  {journal}
  {Scientific Reports}\ }\textbf {\bibinfo {volume} {7}},\ \bibinfo {pages}
  {1122} (\bibinfo {year} {2017})}\BibitemShut {NoStop}%
\bibitem [{\citenamefont {Lucchese}\ \emph {et~al.}(2010)\citenamefont
  {Lucchese}, \citenamefont {Stavale}, \citenamefont {Ferreira}, \citenamefont
  {Vilani}, \citenamefont {Moutinho}, \citenamefont {B.Capaz}, \citenamefont
  {Achete},\ and\ \citenamefont {A.Jorio}}]{lucchese10}%
  \BibitemOpen
  \bibfield  {author} {\bibinfo {author} {\bibfnamefont {M.}~\bibnamefont
  {Lucchese}}, \bibinfo {author} {\bibfnamefont {F.}~\bibnamefont {Stavale}},
  \bibinfo {author} {\bibfnamefont {E.~M.}\ \bibnamefont {Ferreira}}, \bibinfo
  {author} {\bibfnamefont {C.}~\bibnamefont {Vilani}}, \bibinfo {author}
  {\bibfnamefont {M.}~\bibnamefont {Moutinho}}, \bibinfo {author}
  {\bibfnamefont {R.}~\bibnamefont {B.Capaz}}, \bibinfo {author} {\bibfnamefont
  {C.}~\bibnamefont {Achete}}, \ and\ \bibinfo {author} {\bibnamefont
  {A.Jorio}},\ }\href@noop {} {\bibfield  {journal} {\bibinfo  {journal}
  {Carbon}\ }\textbf {\bibinfo {volume} {48}},\ \bibinfo {pages} {1592}
  (\bibinfo {year} {2010})}\BibitemShut {NoStop}%
\bibitem [{\citenamefont {Mukherjee}\ and\ \citenamefont
  {Kaloni}(2012)}]{SMTK2012}%
  \BibitemOpen
  \bibfield  {author} {\bibinfo {author} {\bibfnamefont {S.}~\bibnamefont
  {Mukherjee}}\ and\ \bibinfo {author} {\bibfnamefont {T.~P.}\ \bibnamefont
  {Kaloni}},\ }\href@noop {} {\bibfield  {journal} {\bibinfo  {journal}
  {Journal of Nanoparticles Research}\ }\textbf {\bibinfo {volume} {14}},\
  \bibinfo {pages} {1059} (\bibinfo {year} {2012})}\BibitemShut {NoStop}%
\bibitem [{\citenamefont {D'Souza}\ and\ \citenamefont
  {Mukherjee}(2015)}]{rdsm15}%
  \BibitemOpen
  \bibfield  {author} {\bibinfo {author} {\bibfnamefont {R.}~\bibnamefont
  {D'Souza}}\ and\ \bibinfo {author} {\bibfnamefont {S.}~\bibnamefont
  {Mukherjee}},\ }\href@noop {} {\bibfield  {journal} {\bibinfo  {journal}
  {Physica E}\ }\textbf {\bibinfo {volume} {69}},\ \bibinfo {pages} {138}
  (\bibinfo {year} {2015})}\BibitemShut {NoStop}%
\bibitem [{\citenamefont {D'Souza}, \citenamefont {Mukherjee},\ and\
  \citenamefont {Saha-Dasgupta}(2017)}]{rst17}%
  \BibitemOpen
  \bibfield  {author} {\bibinfo {author} {\bibfnamefont {R.}~\bibnamefont
  {D'Souza}}, \bibinfo {author} {\bibfnamefont {S.}~\bibnamefont {Mukherjee}},
  \ and\ \bibinfo {author} {\bibfnamefont {T.}~\bibnamefont {Saha-Dasgupta}},\
  }\href@noop {} {\bibfield  {journal} {\bibinfo  {journal} {Journal of Alloys
  and Compounds}\ }\textbf {\bibinfo {volume} {708}},\ \bibinfo {pages} {437}
  (\bibinfo {year} {2017})}\BibitemShut {NoStop}%
\bibitem [{\citenamefont {Bernardi}, \citenamefont {Palummo},\ and\
  \citenamefont {Grossman}(2012)}]{grossman12}%
  \BibitemOpen
  \bibfield  {author} {\bibinfo {author} {\bibfnamefont {M.}~\bibnamefont
  {Bernardi}}, \bibinfo {author} {\bibfnamefont {M.}~\bibnamefont {Palummo}}, \
  and\ \bibinfo {author} {\bibfnamefont {J.}~\bibnamefont {Grossman}},\
  }\href@noop {} {\bibfield  {journal} {\bibinfo  {journal} {Phys. Rev. Lett}\
  }\textbf {\bibinfo {volume} {108}},\ \bibinfo {pages} {226805} (\bibinfo
  {year} {2012})}\BibitemShut {NoStop}%
\bibitem [{\citenamefont {Nika}\ \emph
  {et~al.}(2009{\natexlab{b}})\citenamefont {Nika}, \citenamefont {Ghosh},
  \citenamefont {Pokatilov},\ and\ \citenamefont {Balandin}}]{nika09}%
  \BibitemOpen
  \bibfield  {author} {\bibinfo {author} {\bibfnamefont {D.~L.}\ \bibnamefont
  {Nika}}, \bibinfo {author} {\bibfnamefont {S.}~\bibnamefont {Ghosh}},
  \bibinfo {author} {\bibfnamefont {E.~P.}\ \bibnamefont {Pokatilov}}, \ and\
  \bibinfo {author} {\bibfnamefont {A.~A.}\ \bibnamefont {Balandin}},\
  }\href@noop {} {\bibfield  {journal} {\bibinfo  {journal} {Appl. Phys.
  Lett.}\ }\textbf {\bibinfo {volume} {94}},\ \bibinfo {pages} {203103}
  (\bibinfo {year} {2009}{\natexlab{b}})}\BibitemShut {NoStop}%
\bibitem [{\citenamefont {Cutler}\ and\ \citenamefont {Mott}(1969)}]{mott}%
  \BibitemOpen
  \bibfield  {author} {\bibinfo {author} {\bibfnamefont {M.}~\bibnamefont
  {Cutler}}\ and\ \bibinfo {author} {\bibfnamefont {N.~F.}\ \bibnamefont
  {Mott}},\ }\href@noop {} {\bibfield  {journal} {\bibinfo  {journal} {Physical
  Review}\ }\textbf {\bibinfo {volume} {181}},\ \bibinfo {pages} {1336}
  (\bibinfo {year} {1969})}\BibitemShut {NoStop}%
\end{thebibliography}

%
\end{document}